\documentclass[usenatbib]{mn2e}

\usepackage{amsmath, mathrsfs, amssymb}
\usepackage{graphicx}
\usepackage{rotating}
\usepackage{longtable,lscape}
\usepackage{morefloats}
\usepackage{color}
\usepackage[usenames,dvipsnames,svgnames,table]{xcolor}
\usepackage{enumitem}
\usepackage{caption}
\usepackage{verbatim}
\usepackage{color}

\newcommand{\del}{\partial}

\newcommand\araa{{ARA\&A}}%
\newcommand\apj{{ApJ}}%
\newcommand\apjl{{ApJ}}%
\newcommand\apjs{{ApJS}}%
\newcommand\aap{{A\&A}}%
\newcommand\mnras{{MNRAS}}%
\newcommand\na{{New A}}%
\newcommand\pre{{Phys.~Rev.~E}}%
\title[Accretion disc dynamos]{Accretion disc dynamo activity in local simulations spanning weak-to-strong net vertical magnetic flux regimes}

\author[Salvesen et al.]{Greg~Salvesen$^{1,2}\thanks{E-mail: salvesen@colorado.edu}\thanks{NASA Earth and Space Science Graduate Fellow.}$, Jacob~B.~Simon$^{1,3}$\thanks{Sagan Fellow.}, Philip~J.~Armitage$^{1,2}$, \& \newauthor Mitchell~C.~Begelman$^{1,2}$ \\
$^{1}${JILA, University of Colorado and National Institute of Standards and Technology, 440 UCB, Boulder, CO 80309-0440, USA.} \\
$^{2}${Department of Astrophysical and Planetary Sciences, University of Colorado, 391 UCB, Boulder, CO 80309-0391, USA.} \\
$^{3}${Department of Space Studies, Southwest Research Institute, Boulder, CO 80302, USA.}}

\begin{document}
\label{firstpage}
\maketitle

\begin{abstract}
Strongly magnetized accretion discs around black holes have attractive features that may explain enigmatic aspects of X-ray binary behaviour. The structure and evolution of these discs are governed by a dynamo-like mechanism, which channels part of the accretion power liberated by the magnetorotational instability (MRI) into an ordered toroidal magnetic field.  To study dynamo activity, we performed three-dimensional, stratified, isothermal, ideal magnetohydrodynamic shearing box simulations.  The strength of the self-sustained toroidal magnetic field depends on the net vertical magnetic flux, which we vary across almost the entire range over which the MRI is linearly unstable. We quantify disc structure and dynamo properties as a function of the initial ratio of mid-plane gas pressure to vertical magnetic field pressure, $\beta_0^{\rm mid} = p_{\rm gas} / p_B$. For $10^5 \geq \beta_0^{\rm mid} \geq 10$ the effective $\alpha$-viscosity parameter scales as a power-law. Dynamo activity persists up to and including $\beta_0^{\rm mid} = 10^2$, at which point the entire vertical column of the disc is magnetic pressure-dominated. Still stronger fields result in a highly inhomogeneous disc structure, with large density fluctuations. We show that the turbulent steady state $\beta^{\rm mid}$ in our simulations is well-matched by the analytic model of \citet{Begelman2015} describing the creation and buoyant escape of toroidal field, while the vertical structure of the disc can be broadly reproduced using this model.  Finally, we discuss the implications of our results for observed properties of X-ray binaries.
\end{abstract}


\section{Introduction}
\label{sec:intro}
The magnetorotational instability \citep[MRI;][]{BalbusHawley1991} is a well-understood mechanism for generating turbulence and angular momentum transport in accretion discs. In the limit where the disc has no net magnetic field, the MRI acts as a true dynamo in which the majority of the magnetic energy resides in large-scale toroidal magnetic fields. Local magnetohydrodynamic (MHD) simulations with zero net vertical flux show that the rate of angular momentum transport, parametrized by the effective $\alpha$-viscosity \citep{ShakuraSunyaev1973}, is then of the order of $\alpha \sim 0.01$ \citep{Davis2010}.

The zero net field limit of the MRI provides an answer to the fundamental question of why discs accrete, but it may not be the relevant regime for most astrophysical systems. The rate of angular momentum transport increases with net magnetic flux \citep{Hawley1995} above a threshold --- an initial mid-plane ratio of gas to vertical magnetic field pressure $\beta_0^{\rm mid} \sim 10^5$ --- that is quite low. At much stronger field strength, $\beta_0^{\rm mid} \sim 10^2$, this turbulent transport can be substantial with $\alpha \sim 1$ \citep{BaiStone2013}. The importance of magnetic pressure also increases with net flux. In simulations with weak net flux, magnetic pressure dominates only in a coronal region \citep{MillerStone2000}, whereas fields stronger than $\beta_0^{\rm mid} \sim 10^3$ lead to suprathermal toroidal fields in the disc mid-plane \citep{BaiStone2013} and qualitative changes to the disc structure. Finally, disc winds appear to accompany the MRI whenever a dynamically significant net field is present \citep{SuzukiInutsuka2009,Fromang2013,BaiStone2013}.

Except in special circumstances \citep[such as near magnetospheres, or when discs in binaries are threaded by a secondary star's magnetic field;][]{MeyerHofmeister1996}, determining the net field from first principles is a hard problem. Depending on circumstances, a local net field may be the remnant of that present when the disc formed \citep{SikoraBegelman2013}, the product of dynamo action from zero net field conditions \citep{Beckwith2011}, or the result of a competition between advection and diffusion processes \citep{Lubow1994}. None of these processes is fully understood, but it is plausible that they could typically lead to net fields stronger than the very low threshold for changes to $\alpha$. At a more phenomenological level, disc models that invoke strong fields show promise for modeling observed disc properties that are otherwise inexplicable. The structure of magnetically dominated discs is stable against thermal/viscous instability \citep{BegelmanPringle2007}, and less susceptible to gravitational fragmentation \citep{Pariev2003,BegelmanPringle2007,Gaburov2012}. Moreover, since the {\em evolution} of the net field occurs on a time scale that is intermediate between the dynamical and viscous time scales, it has the potential to act as a slowly varying control parameter in models of X-ray binary state transitions \citep{BegelmanArmitage2014}.

Our goal in this study is to quantify the properties of MRI disc turbulence across almost the entire range of net fluxes that admit linearly unstable MRI modes. Following a long line of prior work, we adopt a local approximation and consider vertically stratified isothermal discs for which the net flux is a conserved quantity. We adopt the basic computational approach of \citet{BaiStone2013}, who performed shearing box simulations with a net vertical magnetic flux that had a ratio of gas to magnetic pressure, $\beta \equiv p_{\rm gas} / p_{B}$, whose initial value at the disc mid-plane ranged from $\beta_{0}^{\rm mid} = 10^{2} - 10^{4}$. We extend their work to stronger fields \citep[$\beta_{0}^{\rm mid} = 10$, previously considered by][]{Lesur2013}, and test the sensitivity of the results to the size of the simulation domain, which affects the variability properties of the zero net flux MRI \citep{Simon2012}. More important than these technical differences, we focus our analysis on the properties of the ``MRI-dynamo" --- the periodic reversals of the large-scale toroidal magnetic field that are characteristic of the weak-field MRI in both the local \citep{Brandenburg1995, Shi2010, Davis2010,Simon2011, Simon2012} and global limits \citep{ONeill2011, Beckwith2011}. If the MRI-dynamo persists at high magnetizations, the Poynting flux associated with the periodic expulsion of magnetic field from the disc could play a dominant role in the energetics of the disc atmosphere \citep{Begelman2015}. Whether this occurs is unclear. \citet{JohansenLevin2008}, using short duration simulations, found apparently stable highly magnetized disc structures, while \citet{BaiStone2013} found that the MRI-dynamo petered out once the net field reached $\beta_{0}^{\rm mid} = 10^{2}$. Perhaps as a consequence of our larger domain size, we find instead that MRI-dynamo activity can be present even when the toroidal magnetic field is suprathermal throughout the disc. We also find that the strongest net fields result in the formation of a highly inhomogeneous disc, and we show that the structure of the simulated discs is well approximated by a simple analytic model of buoyant toroidal field escape.

The outline of the paper is as follows. After describing our simulations (\S \ref{sec:sims}), we characterize the properties of MRI turbulence (\S \ref{sec:MRIturb}) and the MRI-dynamo (\S \ref{sec:MRIdynamo}) as a function of net vertical magnetic flux.  We discuss our results in the context of X-ray binary phenomenology and strongly magnetized accretion discs (\S \ref{sec:disc}), followed by a summary and conclusions (\S \ref{sec:sumconc}).

\section{Numerical Simulations}
\label{sec:sims}
We use the \texttt{Athena} code to solve the equations of ideal MHD in the shearing box approximation.  We refer the reader to \citet{GardinerStone2005, GardinerStone2008} for descriptions of the \texttt{Athena} algorithms and to \citet{Stone2008} for descriptions of their implementation.

\subsection{Shearing Box Simulations}
\label{sec:shbox}
The shearing box \citep{GoldreichLyndenBell1965} models a relatively small patch of a differentially rotating fluid by expanding the equations of motion in a locally co-rotating frame.  To apply the shearing box geometry to an accretion disc \citep{Hawley1995, Brandenburg1995}, we go from a cylindrical frame $\left( R, \phi, z \right)$ into a local Cartesian frame $\left( x, y, z \right)$ using the coordinate transformations: $x = R - R_{0}$, $y = R_{0} \phi$, $z = z$, where $R_{0}$ is the reference radius corresponding to the center of the shearing box and co-rotating with the disc at angular frequency $\mathbf{\Omega} = \Omega \mathbf{\hat{k}}$.  For a shearing box with vertical density stratification, the equations of compressible, isothermal, ideal MHD in conservative form with unit vectors $\mathbf{\hat{i}}, \mathbf{\hat{j}}, \mathbf{\hat{k}}$ are \citep{Hawley1995, Stone1996},
\begin{align}
\frac{\del{\rho}}{\del{t}} = &- \mathbf{\nabla} \cdot \left( \rho \mathbf{v} \right) \label{eqn:mass} \\
\frac{\del{\left( \rho \mathbf{v} \right)}}{\del{t}} = &- \mathbf{\nabla} \cdot \left[ \rho \mathbf{v} \mathbf{v} - \mathbf{B} \mathbf{B} + \left( p_{\rm gas} + \frac{B^{2}}{2} \right) \mathbf{I} \right] \nonumber \\
&+ 2 q \rho \Omega^{2} x \mathbf{\hat{i}} - \rho \Omega^{2} z \mathbf{\hat{k}} - 2 \mathbf{\Omega} \times \left( \rho \mathbf{v} \right) \label{eqn:momentum} \\
\frac{\del{\mathbf{B}}}{\del{t}} = &- \mathbf{\nabla} \cdot \left( \mathbf{v} \mathbf{B} - \mathbf{B} \mathbf{v}\right). \label{eqn:magflux}
\end{align}
Here $\rho$ is the gas density, $p_{\rm gas}$ is the gas pressure, $\mathbf{v}$ is the velocity, $\rho \mathbf{v}$ is the momentum density, and $\mathbf{B}$ is the magnetic field, which in our convention absorbs a factor of $\mu / \left( \sqrt{4 \pi} \right)$, where $\mu = 1$ is the adopted magnetic permeability. $\mathbf{I}$ is the identity matrix that acts on the total pressure $p_{\rm gas} + B^{2} / 2$.  We adopt an isothermal equation of state, $p_{\rm gas} = \rho c_{\rm s}^{2}$, where $c_{\rm s}$ is the sound speed.  The shear parameter is defined as, $q = - d {\rm ln} \left( \Omega \right) / d {\rm ln} \left( R \right)$, and we choose the value $q = 3 / 2$ corresponding to a Keplerian accretion disc.

There are various options for numerically integrating Equations \ref{eqn:mass}-\ref{eqn:magflux} within \texttt{Athena}. We use the shearing box implementation of \citet{StoneGardiner2010}, with an orbital advection scheme to separately evolve the background shear flow, $\mathbf{v}_{\rm sh} = -q \Omega x \mathbf{\hat{j}}$, and the velocity fluctuations, $\mathbf{v}^{\prime} = \mathbf{v} - \mathbf{v}_{\rm sh}$. Using orbital advection gives a substantial improvement in computational performance and solution accuracy. We use the Harten-Lax-van Leer discontinuities (HLLD) Riemann solver \citep{MiyoshiKusano2005, Mignone2007} (see, e.g., \citet{Salvesen2014} for a comparison of solvers in \texttt{Athena}). Radial ($x$), toroidal ($y$), and vertical ($z$) boundary conditions are shearing periodic, strictly periodic, and outflowing, respectively \citep{Hawley1995, Simon2011}.

The shearing box has well-known limitations, including the neglect of curvature terms and radial gradients, and symmetry between positive and negative $x$ that leaves the location of the central object undefined. Two limitations are of particular concern for our application. First, the shearing box is not an ideal setup for representing disc winds, which are present in all net flux MRI simulations. Although the shearing box effective potential is locally the same as for a global \citep{BlandfordPayne1982} wind, the derived solutions can have an unphysical geometry and additionally depend in detail on the vertical boundary conditions \citep{BaiStone2013,Fromang2013}. Winds are not the focus of this paper, but they are present and we cannot readily quantify the extent to which uncertainties in the strength of disc winds propagate to other aspects of the solution. Second is the issue of domain size. The effective turbulent $\alpha$ measured in local MRI simulations converges at modest domain size, but other properties of interest (such as the level of variability) continue to vary even in spatially extended shearing boxes \citep{Simon2012}. There is no way to internally assess whether an unconverged solution in a small box is better or worse than a converged solution in a box so large that neglect of curvature terms (for some assumed disc thickness) is formally unjustified. Ultimately, global simulations are needed to avoid these limitations that are inherent to the shearing box.

\subsection{Initial Setup and Parameters}
\label{sec:ICs}

\begin{table*}
\addtolength{\tabcolsep}{-0pt}
\centering
\begin{tabular}{c c c c c c c c c c}
\hline
\hline
ID & $\left( N_{x}, N_{y}, N_{z} \right)$ & Gris Res. & MRI Res. & $\beta_{0}^{\rm mid}$ & $B_{z}$ & $\langle Q_{x}^{\rm mid} \rangle_{t}$ & $\langle Q_{y}^{\rm mid} \rangle_{t}$ & $\langle Q_{z}^{\rm mid} \rangle_{t}$ & Initial Field \\
 & [$N_{\rm zones}$] & [$N_{\rm zones} / H$] & [$\lambda_{\rm fg} / \Delta x_{i}$] & & & & & & Configuration \\
\hline
ZNVF & (480, 960, 480) & 48 & 15 & $413$ & 0 & $18.1(8)$ & $47(2)$ & $11.2(5)$ & Flux Tube \\
NVF-$\beta 5$ & (360, 720, 360) & 36 & 1.0 & $10^{5}$ & 4.47{\rm e}-3 & $17.3(6)$ & $46(2)$ & $10.9(3)$ & $B_{z}$+Sinusoid \\
NVF-$\beta 4$ & (360, 720, 360) & 36 & 3.3 & $10^{4}$ & 1.41{\rm e}-2 & $29(1)$ & $70(3)$ & $19.7(8)$ & $B_{z}$+Sinusoid \\
NVF-$\beta 3$ & (240, 480, 240) & 24 & 7.0 & $10^{3}$ & 4.47{\rm e}-2 & $6(2){\rm e}1$ & $1.4(3){\rm e}2$ & $4(1){\rm e}1$ & $B_{z}$+Sinusoid \\
NVF-$\beta 2$ & (240, 480, 240) & 24 & 22 & $10^{2}$ & 1.41{\rm e}-1 & $1.2(2){\rm e}2$ & $3.3(7){\rm e}2$ & $9(1){\rm e}1$ & $B_{z}$+Sinusoid \\
NVF-$\beta 1$ & (180, 360, 180) & 18 & 52 & $10^{1}$ & 4.47{\rm e}-1 &  $1.1(1){\rm e2}$ & $2.4(4){\rm e}2$ & $1.2(1){\rm e}2$ & $B_{z}$+Sinusoid \\
\hline
\end{tabular}
\caption{Basic information for the suite of shearing box simulations.  From {\it left} to {\it right} the columns are: simulation identification label, total number of grid zones in each dimension, grid resolution (applies to all dimensions), number of grid zones spanning the fastest growing MRI wavelength (applies to all dimensions), initial ratio of gas pressure-to-magnetic pressure at the disc mid-plane, net vertical magnetic flux density (code units), saturated state time-averaged $Q_{x}$, $Q_{y}$, $Q_{z}$ (see Equation \ref{eqn:Q}) at the disc mid-plane, and the initial magnetic field configuration.  Parentheses indicate the $\pm 1 \sigma$ range on the last digit.  The domain size for all simulations is $\left( L_{x}, L_{y}, L_{z} \right) = \left( 10H_{0}, 20H_{0}, 10H_{0} \right)$ and all simulations begin at time $t = 0$ and terminate at time $t_{\rm f} = 225$ orbits, while the time-averaging is done over the domain $\left[ t_{\rm i}, t_{\rm f} \right] = \left[ 25, 225 \right]$ orbits.}
\label{tab:siminfo}
\end{table*}

Our initial setup is based on \citet{Stone1996} and closely follows that of \citet{Simon2012}.  We initialize all simulations with a vertically stratified density profile according to isothermal hydrostatic equilibrium,
\begin{equation}
\rho_{0} \left(x, y, z \right) = \rho_{0}^{\rm mid} {\rm exp} \left( \frac{- z^{2}}{2 H_{0}^{2}} \right),
\end{equation}
where $\rho_{0}^{\rm mid}$ is the initial (i.e., $t = 0$) gas density at the disc mid-plane (i.e., $z = 0$) and $H_{0}$ is the initial gas density scale height of the disc\footnote{We note that \citet{Simon2012} define $H_{0} = \sqrt{2} c_{\rm s} / \Omega$; therefore, one must keep this in mind if comparing our results to theirs.},
\begin{equation}
H_{0} = \frac{c_{\rm s}}{\Omega}.
\end{equation}
In code units the initial parameter choices are $\Omega = 1$, $\rho_{0}^{\rm mid} = 1$, and $c_{\rm s} = 1$, which correspond to $H_{0} = 1$ and initial disc mid-plane gas pressure $p_{{\rm gas}, 0}^{\rm mid} = 1$.

We run shearing box simulations with and without an imposed net vertical magnetic flux.  The magnetization of the gas is parametrized by the ratio of gas pressure-to-magnetic pressure, defined as,
\begin{equation}
\beta = \frac{p_{\rm gas}}{p_{\rm B}} = \frac{\rho c_{\rm s}^{2}}{B^{2} / 2}, \label{eqn:beta}
\end{equation}
with $\beta_{0}$ being the initial plasma-$\beta$ parameter and $\beta_{0}^{\rm mid}$ being its value at the disc mid-plane.  Our net vertical magnetic flux simulations adopt an initial magnetic field configuration, $\mathbf{B}_{0}$, consisting of a uniform vertical field with an additional sinusoidal component (added to improve numerical stability during the initial transient growth of the MRI),
\begin{align}
B_{x, 0} \left(x, y, z \right) &= 0 \nonumber \\
B_{y, 0} \left(x, y, z \right) &= 0 \nonumber \\
B_{z, 0} \left(x, y, z \right) &= B_{0} \left[ 1 + \frac{1}{2} {\rm sin} \left( \frac{2 \pi x}{L_{x}} \right) \right]. 
\end{align}
Because of the spatial variation of $B_{z,0}$ in the initial conditions, slightly different definitions of the relation between $B_0$ and $\beta_{0}^{\rm mid}$ are possible. We adopt $B_{0} = \sqrt{2 p_{{\rm gas}, 0} / \beta_{0}^{\rm mid}}$.  For these simulations initialized with a net vertical magnetic flux, $\beta_{0}$ refers only to the vertical component.

The only physical parameter we vary in our suite of net vertical magnetic flux simulations is $\beta_{0}^{\rm mid}$ (see Table \ref{tab:siminfo}).  We choose initial net vertical magnetic flux --- NVF prefix in the simulation naming convention --- values corresponding to very weak (NVF-$\beta 5$; $\beta_{0}^{\rm mid} = 10^{5}$), weak (NVF-$\beta 4$; $\beta_{0}^{\rm mid} = 10^{4}$), moderate (NVF-$\beta 3$; $\beta_{0}^{\rm mid} = 10^{3}$), strong (NVF-$\beta 2$; $\beta_{0}^{\rm mid} = 10^{2}$), and very strong (NVF-$\beta 1$; $\beta_{0}^{\rm mid} = 10^{1}$) magnetization levels.

For comparison we also run a zero net vertical magnetic flux simulation (ZNVF; see Table \ref{tab:siminfo}), which adopts an identical initial magnetic field configuration to that of \citet{Simon2012}, based on the twisted toroidal flux tube setup of \citet{Hirose2006}.  As was done in \citet{Simon2012}, the initial toroidal and poloidal plasma-$\beta$ parameters for this simulation are $\beta_{y, 0} = 100$ and $\beta_{p, 0} = 1600$, respectively, corresponding to $\beta_{0}^{\rm mid} = 413$.

In order to seed the MRI, we populate the grid at time $t = 0$ with perturbations --- drawn randomly from a uniform distribution with zero mean value --- to the gas density and each velocity component.  The maximum density and velocity perturbation amplitudes are $| \delta \rho | / \rho = 0.01$ and $| \delta v_{i} | = \left(0.01 / 5 \right) c_{s}$, respectively \citep[e.g.,][]{Hawley1995}, where $i = \left( x, y, z \right)$ denotes the spatial dimension. 

To prevent time-steps from becoming too small as a result of large Alfv\'{e}n speeds in regions of very low gas density, we enforce a gas density floor of $\rho_{\rm floor} = 10^{-4} \rho_{0}^{\rm mid}$.  For all simulations, the vertical profiles of horizontally-averaged gas density remain well above this lower limit for all heights above/below the disc mid-plane.

For all simulations, we choose a domain size of $\left(L_{x}, L_{y}, L_{z} \right) = \left( 10 H_{0}, 20 H_{0}, 10 H_{0} \right)$.  This choice is motivated by the vertically stratified shearing box simulations of \citet{Simon2012}, which provide compelling evidence for non-local, ``mesoscale'' structures (i.e., on scales $\gg H$) contributing to angular momentum transport as the domain size increases \citep[see also][]{GuanGammie2011}. It also allows us to explore the sensitivity of the results to changes in domain size by comparison (where our parameters overlap) with the results of \cite{BaiStone2013}, who used very similar computational methods. As we have already remarked, the neglect of curvature terms in large local domains, such as ours, is only formally justified when modeling very thin discs (with $H / R \sim 0.01$). Explicit comparisons of dynamo behavior in mesoscale and global simulations \citep{Beckwith2011}, however, show that this aspect of the dynamics remains well-modeled locally even when the formal criterion for validity is only marginally satisfied. Simulations NVF-$\beta 5$ and NVF-$\beta 4$ have a spatial resolution of $36~{\rm grid~zones} / H_{0}$ in each dimension, or $\left( N_{x}, N_{y}, N_{z} \right) = \left( 360, 720, 360 \right)$.  Simulations NVF-$\beta 3$ and NVF-$\beta 2$ have a spatial resolution of $24~{\rm grid~zones} / H_{0}$ in each dimension, or $\left( N_{x}, N_{y}, N_{z} \right) = \left( 240, 480, 240 \right)$.  Simulation NVF-$\beta 1$ has a spatial resolution of $18~{\rm grid~zones} / H_{0}$ in each dimension, or $\left( N_{x}, N_{y}, N_{z} \right) = \left( 180, 360, 180 \right)$.  The zero net magnetic flux simulation ZNVF has a spatial resolution of $48~{\rm grid~zones} / H_{0}$ in each dimension, or $\left( N_{x}, N_{y}, N_{z} \right) = \left( 480, 960, 480\right)$.  Table \ref{tab:siminfo} summarizes the suite of shearing box simulations considered in this work.

Ideally, we would like to resolve both the most unstable linear MRI modes in the initial conditions (which can be assessed analytically), and the non-linear properties of turbulence in the saturated state (which can only be assessed empirically). This aspiration is very hard to achieve across the range of net fluxes we consider, because the most unstable linear modes are of much smaller scale in the weak field simulations\footnote{In principle, the weak field simulations could be run in smaller boxes, with higher spatial resolution, but this would introduce different but equally problematic issues of convergence with domain size.}. Considering first the linear criterion, the fastest growing wavelength for the MRI for an {\it unstratified} isothermal Keplerian accretion disc is \citep{Hawley1995},
\begin{equation}
\lambda_{\rm fg} = \frac{8 \pi}{\sqrt{15}} \frac{v_{{\rm A}, 0}}{\Omega},
\end{equation}
where $v_{{\rm A}, 0} = B_{0} / \sqrt{\rho_{0}}$ is the initial Alfv\'{e}n speed.  Table \ref{tab:siminfo} lists the number of grid zones resolving $\lambda_{\rm fg}$ at the disc mid-plane for each of our simulations. Clearly our two weakest net field runs (NVF-$\beta 5$ and NVF-$\beta 4$) fail to properly resolve this mode (the run with {\em zero} net vertical flux is initialized with a moderately strong toroidal field, whose instability is easier to resolve). We can also consider the resolution needed to reproduce the fastest growing modes in {\it stratified} discs. \citet{Latter2010} performed one-dimensional numerical simulations to compute MRI growth rates in this case and suggest that at least $25, 50, 200 {\rm~grid~zones} / H$ are needed for simulations with $\beta_{0}^{\rm mid} = 10^{2}, 10^{3}, 10^{4}$, respectively, in order to resolve the fastest growing linear MRI modes that develop into channel flows. We meet this requirement for NVF-$\beta 2$, but not for the weaker net field cases.

The above discussion indicates that the linear growth of the MRI in the two weakest net field runs cannot be reliably captured by our simulations. Our analyses exclude this phase. All of the runs, however, are unstable and develop into a turbulent saturated state in which the magnetic field is substantially stronger (and hence easier to resolve). Empirically, the resolution of the turbulent steady state can be parametrized by the quality factor \citep{Sano2004},
\begin{equation}
Q_{i} = \frac{\lambda_{{\rm MRI}, i}}{\Delta x_{i}}, \label{eqn:Q}
\end{equation}
where $\lambda_{{\rm MRI}, i} = 2 \pi v_{{\rm A}, i} / \Omega$ is the characteristic MRI wavelength along spatial dimension $i$, $v_{{\rm A}, i} = \sqrt{B_{i}^{2} / \rho}$ is the Alfv\'{e}n speed in direction $i$, and $\Delta x_{i}$ is the size of a grid zone in direction $i$.  \citet{Sano2004} suggest a turbulent steady state time-averaged quality factor $\langle Q_{i} \rangle_{t} \gtrsim 6$, while \citet{Hawley2011} recommend a more stringent $\langle Q_{y} \rangle_{t} \gtrsim 20$ and $\langle Q_{z} \rangle_{t} \gtrsim 10$.  These criteria are good rules of thumb, though clearly the specific $\langle Q_{i} \rangle_{t}$ values needed to properly resolve turbulence are code dependent.  Table \ref{tab:siminfo} lists the horizontally- and time-averaged quality factors evaluated at the disc mid-plane (i.e., where $Q_{i}$ is minimized).  
For all of our simulations, the turbulent steady state is well-resolved by the standards just outlined.

Due to the net vertical magnetic flux we impose, significant mass outflows develop that would deplete the entire disc on timescales shorter than the 200 orbits that we study in the turbulent steady state.  Therefore, after every time step, we maintain a constant mass in the domain by multiplying the gas density in every grid zone by an appropriate common factor \citep{BaiStone2013}.  

\subsection{Notation Convention}
\label{sec:notation}
In the analyses described in subsequent sections, angled brackets surrounding a quantity denote an average.  The subscript on the brackets indicates the dimension being averaged over, where $\langle G \rangle_{t}$, $\langle G \rangle_{xy}$, and $\langle G \rangle_{V}$ indicate a time average, horizontal (i.e., disc-plane) average, and volume average of the quantity $G \left( x, y, z, t \right)$, respectively, given by,
\begin{align}
\langle G \rangle_{t} &= \frac{1}{t_{\rm f} - t_{\rm i}} \int_{t_{\rm i}}^{t_{\rm f}} G dt \\
\langle G \rangle_{xy} &= \frac{1}{L_{x} L_{y}} \int_{-\frac{1}{2}L_{y}}^{+\frac{1}{2}L_{y}} \int_{-\frac{1}{2}L_{x}}^{+\frac{1}{2}L_{x}} G dx dy \\
\langle G \rangle_{V} &= \frac{1}{L_{z}} \int_{-\frac{1}{2}L_{z}}^{+\frac{1}{2}L_{z}} \langle G \rangle_{xy} dz.
\end{align}
Time averages span 200 orbits over the range $\left[ t_{\rm i}, t_{\rm f} \right] = [25, 225]$, where we measure time units in orbits at the center of the shearing box.  Horizontal averages encompass the domain $\left( L_{x}, L_{y} \right) = \left( 10 H_{0}, 20 H_{0} \right)$.  Volume averages encompass the domain $\left( L_{x}, L_{y}, L_{z} \right) = \left( 10 H_{0}, 20 H_{0}, 10 H_{0} \right)$.

Nested angled brackets denote averages over multiple dimensions, where $\langle \langle G \rangle_{V} \rangle_{t}$ means that the quantity $G$ was volume-averaged and then time-averaged.  A quantity with the superscript $G^{\rm mid}$ means that $G$ was evaluated at the disc mid-plane.  A quantity with the subscript $G_{0}$ means that $G$ was evaluated at $t = 0$.  Quoted uncertainty ranges or error bands shown in figures denote the $\pm 1\sigma$ range corresponding only to the dimension indicated by the subscript on the outermost angled brackets.

Finally, we introduce the notations of a bar ($-$) and a hat ($\wedge$) over a parameter that is defined as the ratio of two quantities, such as $\beta \equiv p_{\rm gas} / p_{B}$ (Equation \ref{eqn:beta}) and $\alpha \equiv T_{xy} / p_{\rm gas}$ (Equation \ref{eqn:alpha}).  Using $\beta$ as an example, the bar and hat have the following meanings,
\begin{align}
\overline{\beta} &\equiv \frac{\langle p_{\rm gas} \rangle_{V}}{\langle p_{B}\rangle_{V}} \label{eqn:beta_bar} \\
\widehat{\beta} &\equiv \frac{\langle p_{\rm gas} \rangle_{xy}}{\langle p_{B}\rangle_{xy}}. \label{eqn:beta_hat}
\end{align}

\section{Basic Properties of MRI Turbulence}
\label{sec:MRIturb}

\begin{figure}
  \includegraphics[width=84mm]{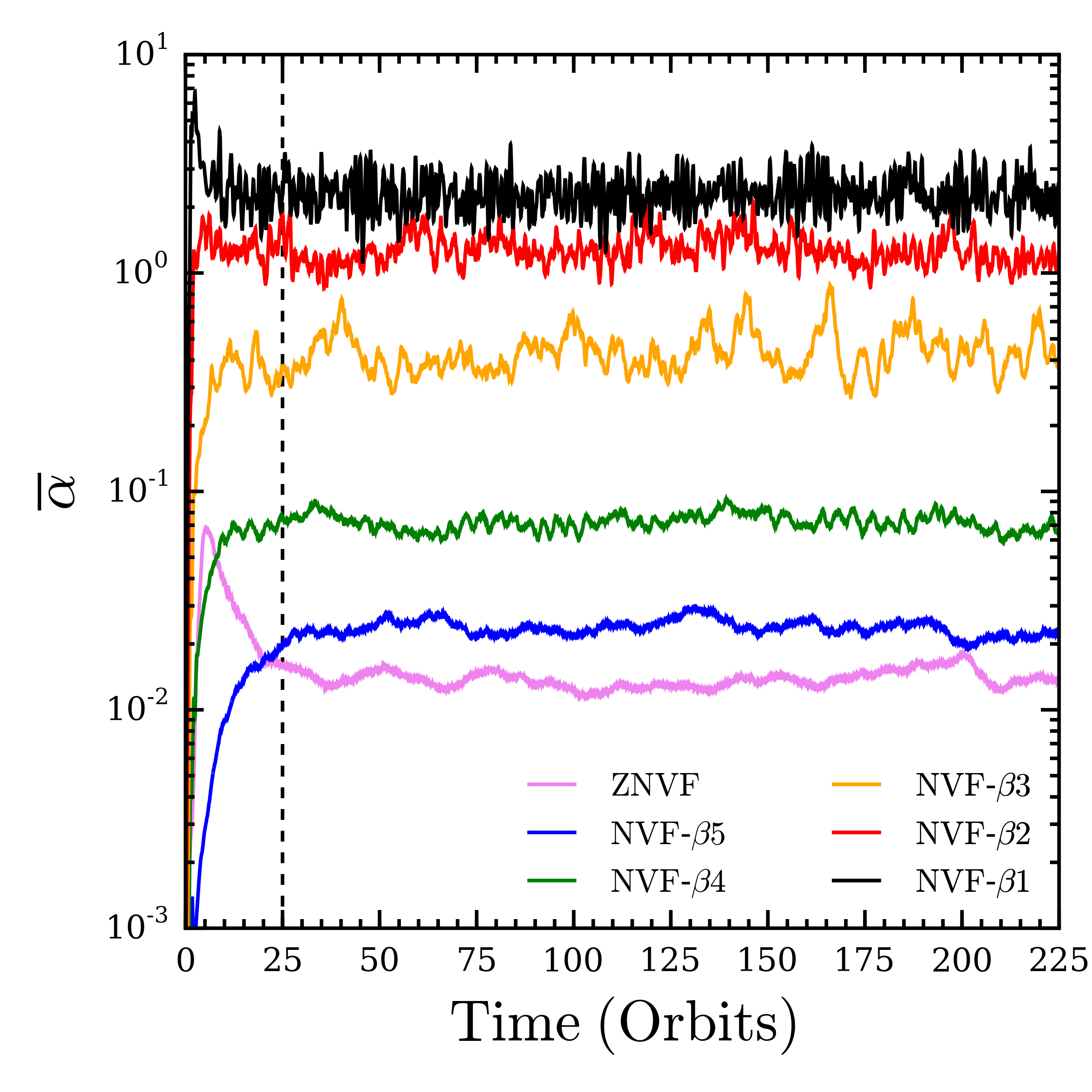}
  \caption{Time evolution of the effective viscosity parameter $\overline{\alpha} = \langle T_{xy} \rangle_{V} / \langle p_{\rm gas} \rangle_{V}$ for each simulation: ZNVF ({\it purple lines}), NVF-$\beta 5$ ({\it blue lines}), NVF-$\beta 4$ ({\it green lines}), NVF-$\beta 3$ ({\it orange lines}), NVF-$\beta 2$ ({\it red lines}), and NVF-$\beta 1$ ({\it black lines}).  The {\it vertical dashed line} at time $t = 25$ orbits marks the start of all time averaging in this work.  The effective viscosity parameter is relatively small for the zero net vertical magnetic flux simulation ZNVF, but increases dramatically with increasing net vertical magnetic flux, even exceeding unity for simulations NVF-$\beta 2$ and NVF-$\beta 1$.}
  \label{fig:alpha_tevol}
\end{figure}

\begin{figure}
  \includegraphics[width=84mm]{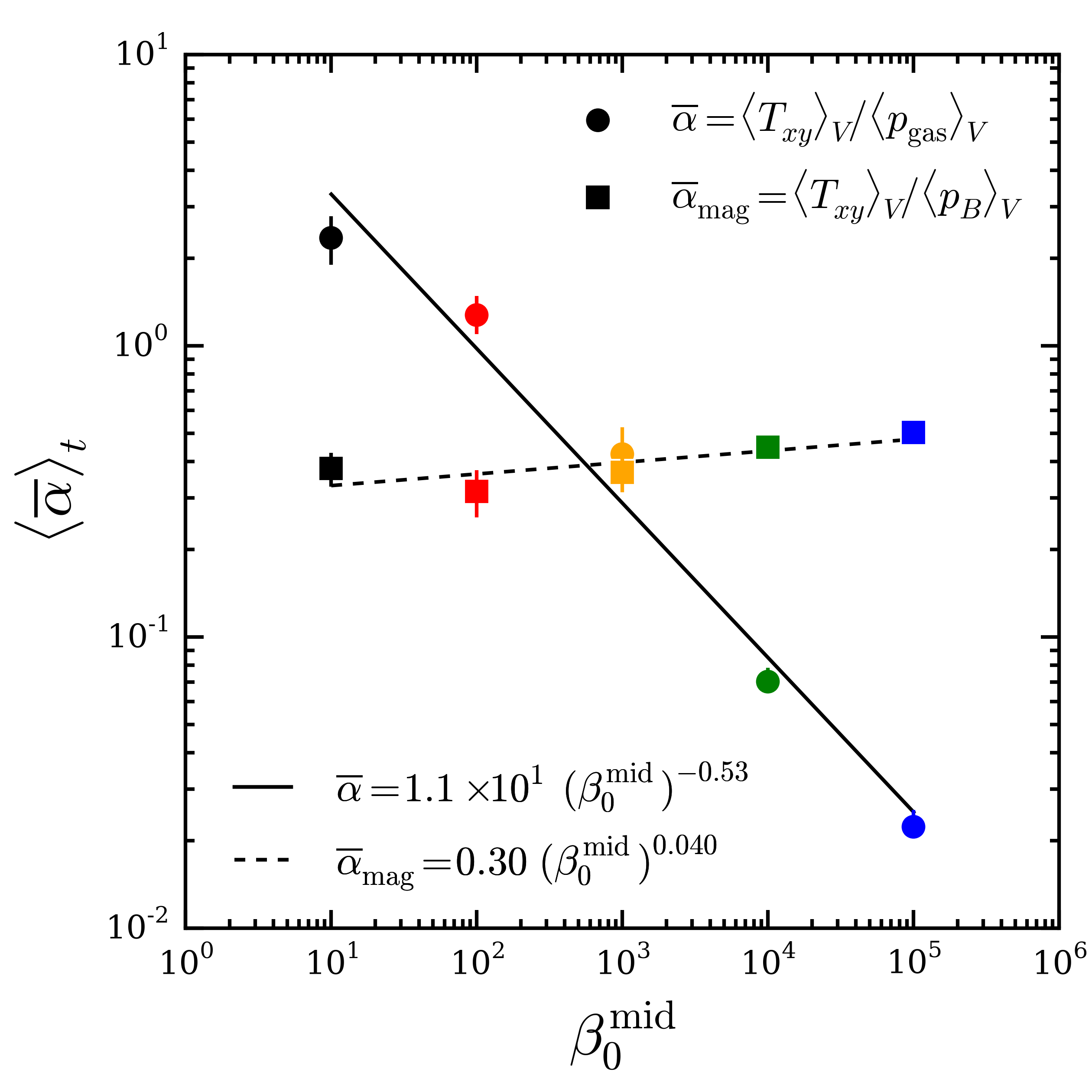}
  \caption{Time-averaged effective $\overline{\alpha}$-viscosity parameter as a function of the initial disc mid-plane ratio of gas pressure-to-magnetic pressure for simulations with a net vertical magnetic flux.  The angular momentum transport rate $\langle \overline{\alpha} \rangle_{t}$ is set by $\beta_{0}^{\rm mid}$, following a power-law relation with the net vertical magnetic flux.  Rather than normalizing the total stress by the gas pressure, $\overline{\alpha} = \langle T_{xy} \rangle_{V} / \langle p_{\rm gas} \rangle_{V}$ ({\it circles}), if we instead normalize by the magnetic pressure, $\overline{\alpha}_{\rm mag} = \langle T_{xy} \rangle_{V} / \langle p_{B} \rangle_{V}$ ({\it squares}), then $\overline{\alpha}_{\rm mag}$ becomes essentially independent of $\beta_{0}^{\rm mid}$.}
  \label{fig:alpha_beta}
\end{figure}

A standard diagnostic of MRI turbulence is the rate of angular momentum transport.  For a Keplerian accretion disc, this is parametrized by the non-dimensional effective $\alpha$-viscosity parameter \citep{ShakuraSunyaev1973},
\begin{align}
\alpha &= \frac{T_{xy}}{p_{\rm gas}} \label{eqn:alpha} \\
\alpha_{\rm Rey} &= \frac{T_{xy, {\rm Rey}}}{p_{\rm gas}} \\
\alpha_{\rm Max} &= \frac{T_{xy, {\rm Max}}}{p_{\rm gas}},
\end{align}
where $T_{xy} = T_{xy, {\rm Rey}} + T_{xy, {\rm Max}}$ is the $x y$ (i.e., $r \phi$) component of the total stress tensor, with Reynolds stress and Maxwell stress components,
\begin{align}
T_{xy, {\rm Rey}} &= \rho v_{x} v_{y}^{\prime} \\
T_{xy, {\rm Max}} &= - B_{x} B_{y}.
\end{align}
For each simulation, Table \ref{tab:MRI_properties} provides values for $\widehat{\beta}$ and $\widehat{\alpha}$, along with their individual components, all evaluated at the disc mid-plane and time-averaged.

\begin{figure*}
    \includegraphics[width=0mm]{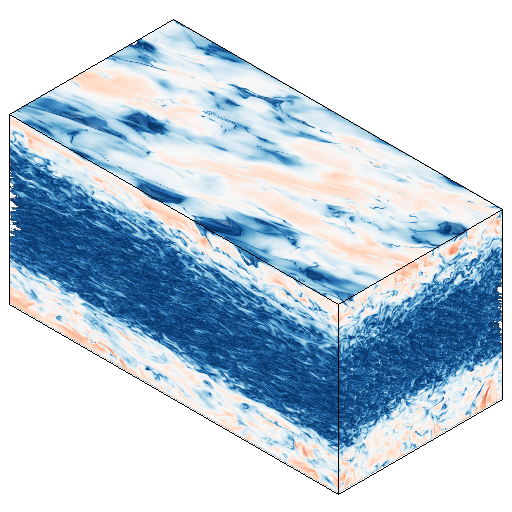}  
    \includegraphics[width=0mm]{fig3a}  
\vspace{-9mm}
    \includegraphics[width=84mm]{fig3a}  
\vspace{-9mm}
    \includegraphics[width=84mm]{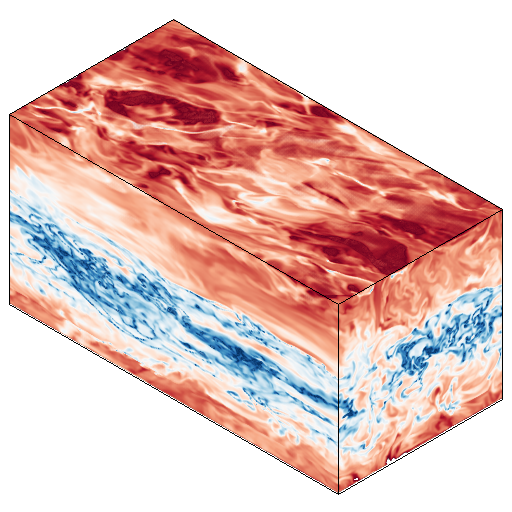}  
\vspace{-9mm}
    \includegraphics[width=84mm]{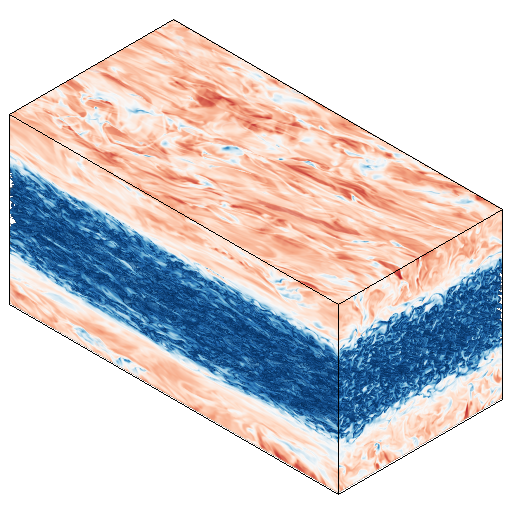}  
\vspace{-9mm}
    \includegraphics[width=84mm]{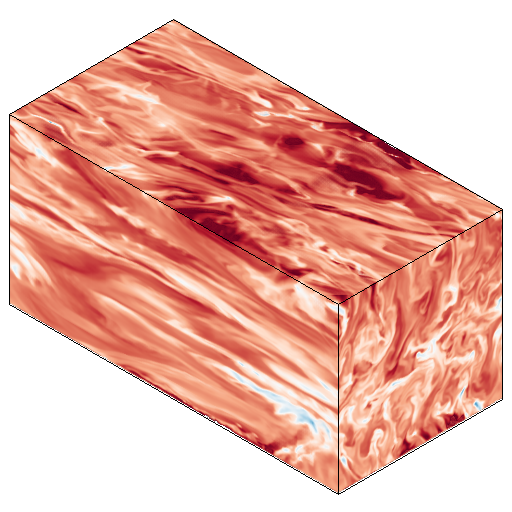}  
\vspace{-9mm}
    \includegraphics[width=84mm]{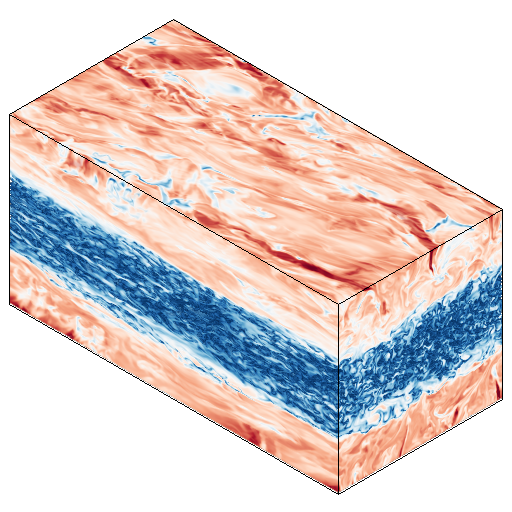}  
\vspace{+3mm}
    \includegraphics[width=84mm]{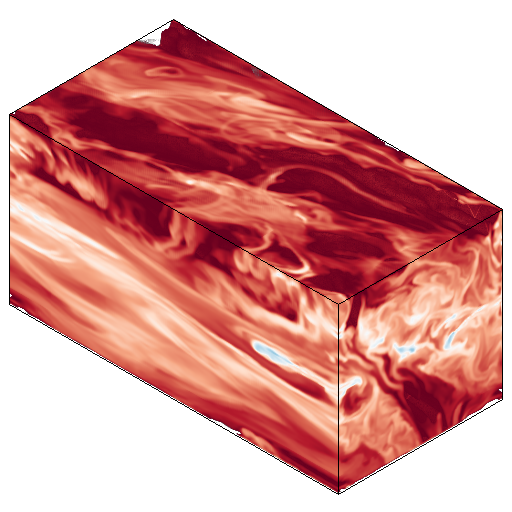}  
\vspace{+3mm}
    \caption{Volume renderings of the ratio of gas pressure-to-magnetic pressure, $\beta \equiv p_{\rm gas} / p_{\rm B}$, at time $t = 125$ orbits.  From {\it top} to {\it bottom}, the {\it left column} shows simulations ZNVF, NVF-$\beta 5$, NVF-$\beta 4$ and the {\it right column} shows simulations NVF-$\beta 3$, NVF-$\beta 2$, NVF-$\beta 1$.  {\it Red}, {\it white}, and {\it blue} colours denote regions with $\beta < 1$, $\beta \simeq 1$, and $\beta > 1$, respectively, with the colours being logarithmically rendered over the range ${\rm log_{10}} \left( \beta \right) = \left[ -2, 2 \right]$.  Larger scale turbulent structures develop with decreasing $\beta_{0}^{\rm mid}$.  The full simulation domain is magnetic pressure-dominated for NVF-$\beta 2$ and NVF-$\beta 1$.}
    \label{fig:volrend}
\end{figure*}

\begin{figure}
  \includegraphics[width=84mm]{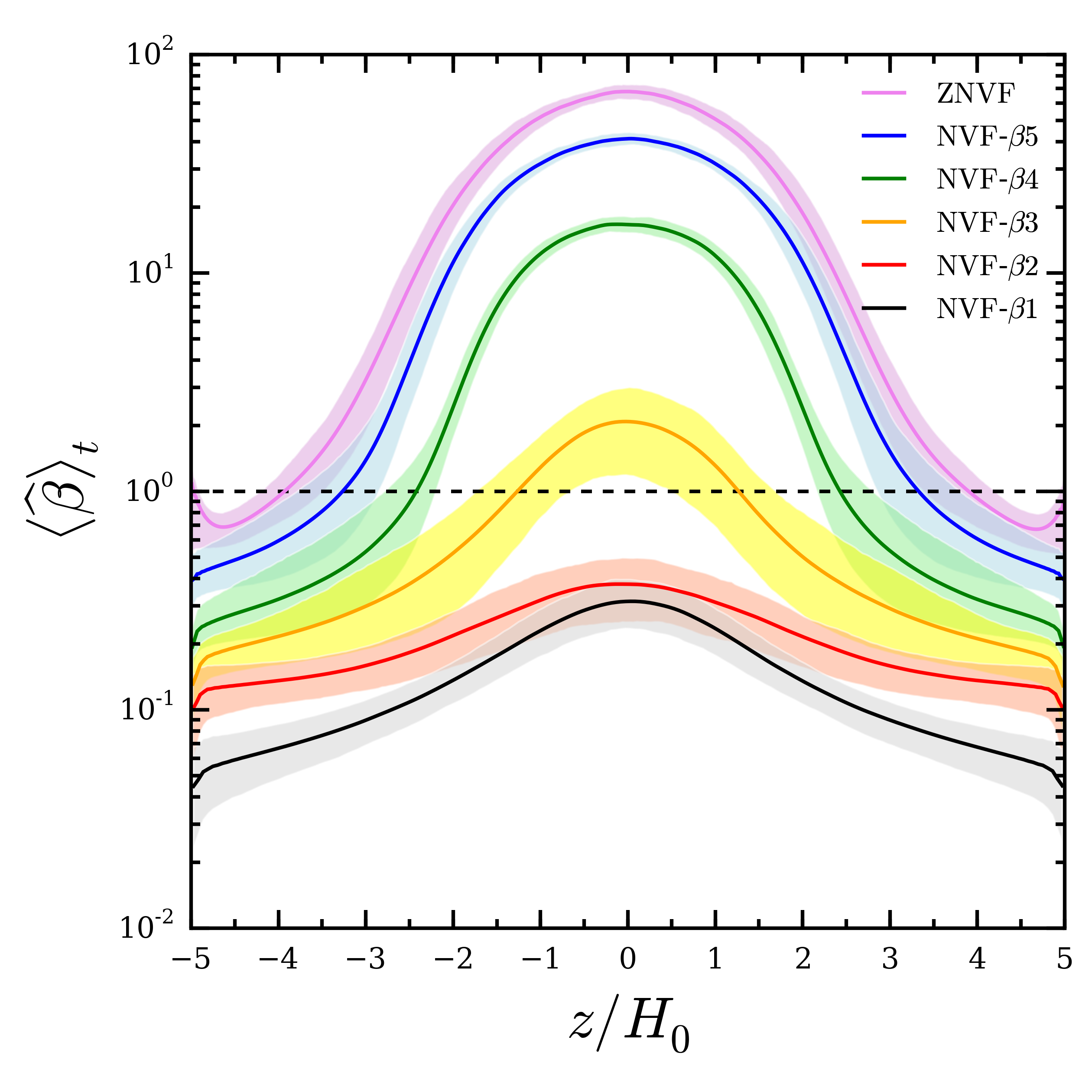}
  \caption{Vertical profiles of the time-averaged $\widehat{\beta}$ (Equation \ref{eqn:beta_hat}) for each simulation (see Figure \ref{fig:alpha_tevol} for line colour conventions).  Coloured bands show the $\pm1 \sigma$ range about the average for the dimension indicated by the outermost angled bracket (time in this case).  The {\it horizontal dashed line} marks equipartition between gas pressure and magnetic pressure.  A magnetic pressure-dominated corona forms for all simulations.  The location of the transition from gas to magnetic pressure-domination (i.e., $\beta = 1$) moves toward the disc mid-plane with increasing net vertical magnetic flux, with the entire domain becoming magnetic pressure-dominated (i.e., $\beta < 1$) for simulations NVF-$\beta 2$ and NVF-$\beta 1$.}
  \label{fig:zpro_beta}
\end{figure}

\begin{figure}
  \includegraphics[width=84mm]{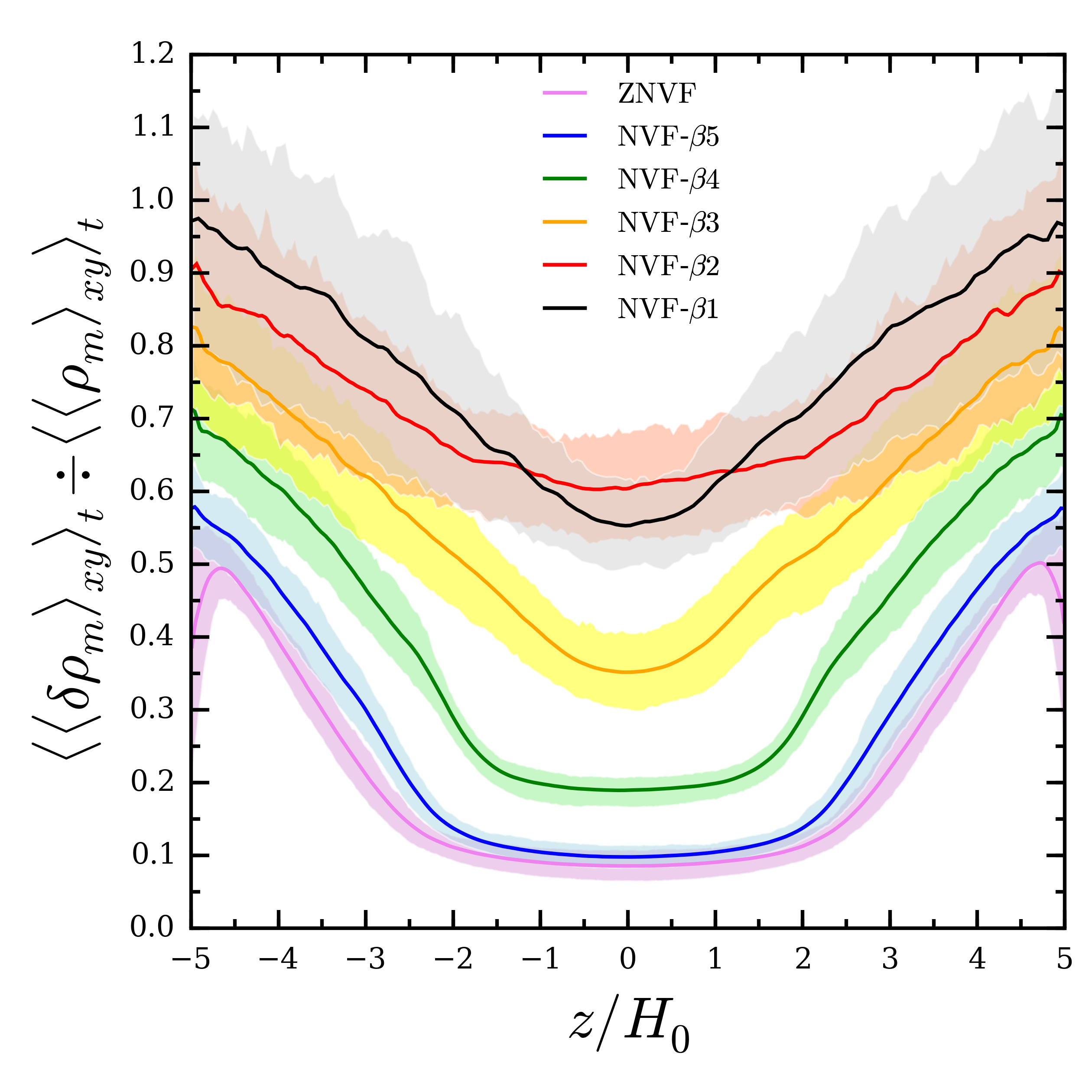}
  \caption{Vertical profiles of the horizontally- and time-averaged gas density fluctuations relative to the horizontally- and time-averaged gas density weighted by mass (see Figure \ref{fig:alpha_tevol} for line colour conventions).  For the zero net vertical magnetic flux simulation ZNVF, the normalized gas density r.m.s. fluctuations are at the $\simeq 10\%$ ($\sim 20-50\%$) level in the disc (corona) regions.  These gas density fluctuations increase with the amount of net vertical magnetic flux, with large variations about the $\simeq 65\%$ ($\sim 75-100\%$) level in the disc (corona) regions for the most strongly magnetized simulations NVF-$\beta 2$ and NVF-$\beta 1$.}
  \label{fig:zpro_densfluc}
\end{figure}

Figure \ref{fig:alpha_tevol} shows the time evolution of $\overline{\alpha}$ for each simulation.  The zero net vertical magnetic flux simulation ZNVF shows a relatively small $\langle \overline{\alpha} \rangle_{t} = 0.015 \pm 0.003$ with little temporal variability, consistent with the results from \citet{Davis2010} and \citet{Simon2012}.  When a net vertical magnetic flux is introduced, the rate of angular momentum transport becomes both enhanced and more variable, achieving $\langle \overline{\alpha} \rangle_{t} = 2.3 \pm 0.5$ for the simulation with the strongest net vertical magnetic flux NVF-$\beta 1$.

Treating the initial value of $\beta$ at the disc mid-plane as a control parameter, Figure \ref{fig:alpha_beta} shows how $\overline{\alpha}$ scales with $\beta_{0}^{\rm mid}$. We find that the scaling is well-fit by a single power-law, which closely matches the expected slope of $-1/2$. Our best fit has the form,
\begin{equation}
\langle \overline{\alpha} \rangle_{t} = 1.1 \times 10^{1} \left( \beta_{0}^{\rm mid} \right)^{-0.53}. \label{eqn:alpha}
\end{equation}
This relation holds over four orders of magnitude in $\beta_{0}^{\rm mid}$ (two orders of magnitude in $B_{z}$), and covers almost the entire range of net fluxes for which the flux (a) boosts the efficiency of transport as compared to zero-net field simulations and (b) allows linearly unstable MRI modes. Alternatively, we can choose to normalize the stress by the magnetic pressure, $\overline{\alpha}_{\rm mag} = \langle T_{xy} \rangle_{V} / \langle p_{B} \rangle_{V}$. With this definition, the effective viscosity becomes essentially independent of $\beta_{0}^{\rm mid}$,
\begin{equation}
\langle \overline{\alpha}_{\rm mag} \rangle_{t} = 0.30 \left( \beta_{0}^{\rm mid} \right)^{-0.040}. \label{eqn:alpha_mag}
\end{equation}
These results are consistent with previous works that find power-law and nearly constant scalings with $\beta$ for $\overline{\alpha}$ and $\overline{\alpha}_{\rm mag}$ \citep{Hawley1995, Blackman2008}. Finally, we can combine Equations \ref{eqn:alpha} and \ref{eqn:alpha_mag} to assess how the disc magnetization (which in the saturated state is dominated by the toroidal component) scales with the imposed net flux. We find,
\begin{equation}
 \langle \overline{\beta} \rangle_{t} \sim 0.03 \left( \beta_{0}^{\rm mid} \right)^{1/2},
\end{equation} 
which implies that the net flux threshold for the disc to become magnetically dominated (in a volume-averaged sense) is at approximately $\beta_{0}^{\rm mid} \sim 10^3$.  As a caveat, we mention that results derived from volume averages of $\alpha$ and $\beta$ will be somewhat dependent on the vertical domain size. For instance, results would be weighted more to the corona for a more vertically extended box.

\begin{figure*}
    \includegraphics[width=84mm]{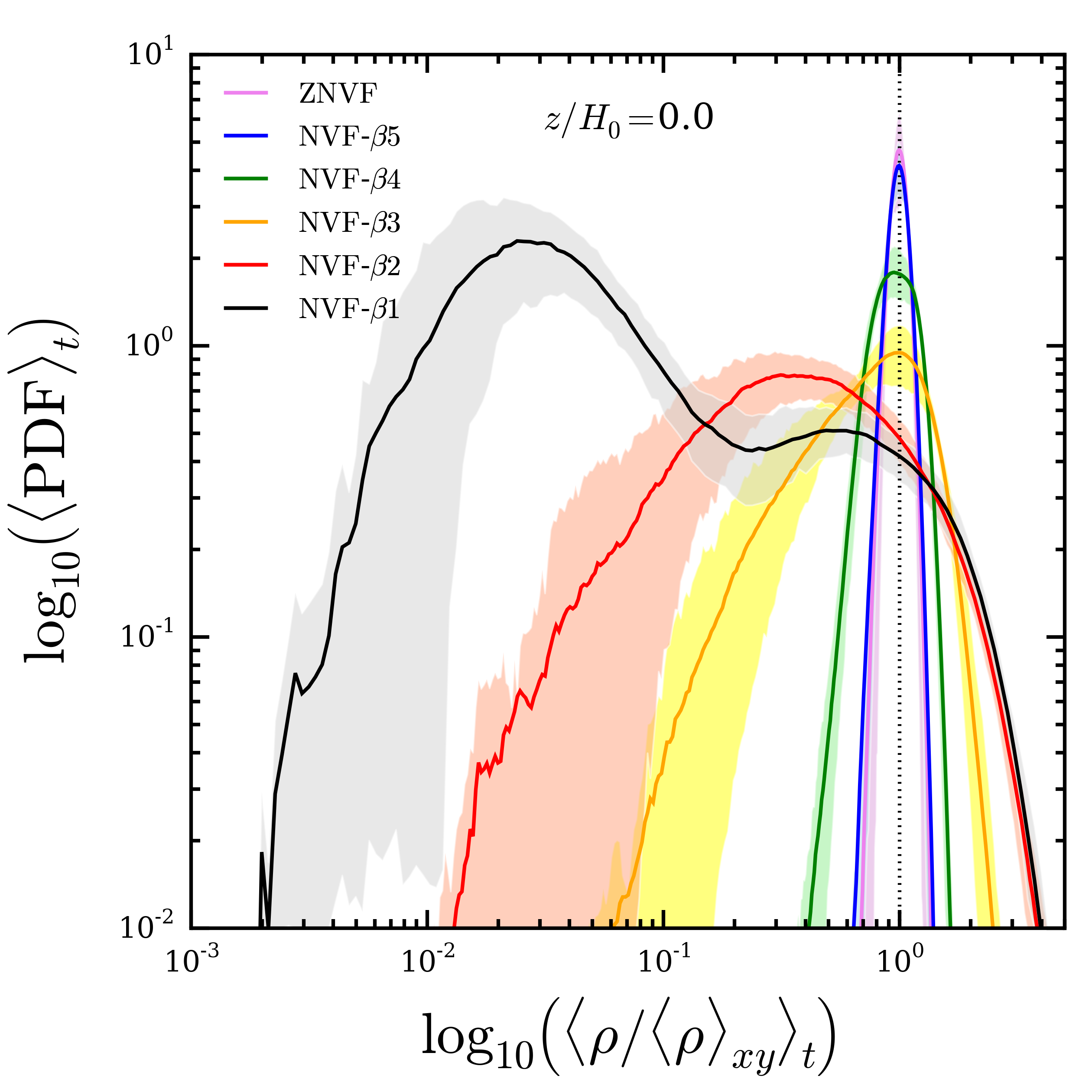}
    \includegraphics[width=84mm]{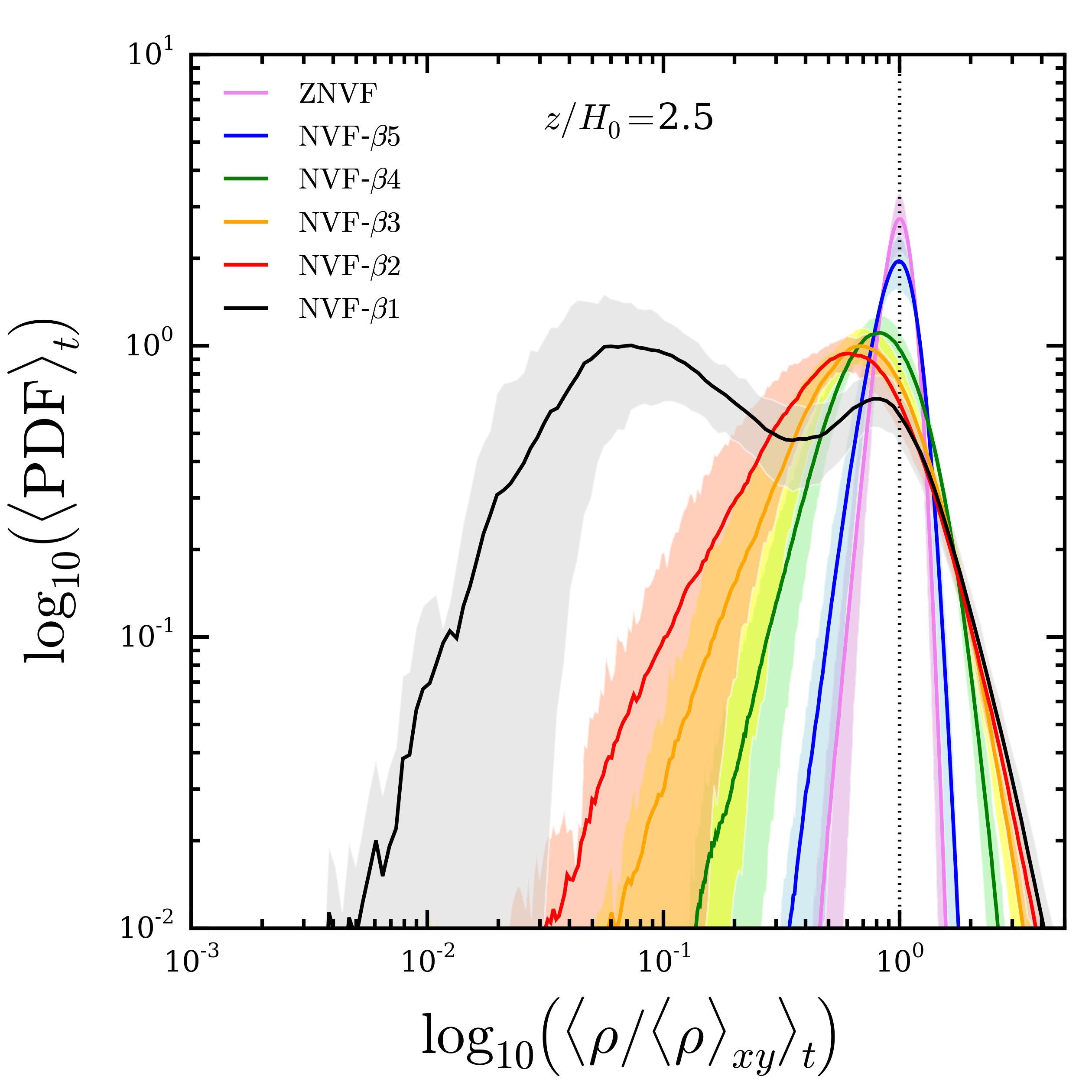}
  \caption{Normalized probability density function of gas density relative to its mean value, evaluated at $z / H_{0} = 0$ ({\it left}) and $z / H_{0} = 2.5$ ({\it right}) for each simulation (see Figure \ref{fig:alpha_tevol} for line colour conventions).  Simulations ZNVF, NVF-$\beta 5$, and NVF-$\beta 4$ have highly homogeneous density structures at the disc mid-plane region, with modest inhomogeneity developing in the coronal regions.  The strongly magnetized simulations NVF-$\beta 3$ and NVF-$\beta 2$ are inhomogeneous throughout the domain, while the very strongly magnetized simulation NVF-$\beta 1$ is highly inhomogeneous.}
  \label{fig:dens_PDF}
\end{figure*}

Figure \ref{fig:volrend} shows volume renderings of $\beta$ at $t = 125$ orbits for each simulation.  The gas is turbulent, with {\it blue} and {\it red} indicating gas pressure-dominated ($\beta > 1$) and magnetic pressure-dominated ($\beta < 1$) regions, respectively, while {\it white} shows regions near equipartition ($\beta \simeq 1$).  As the amount of net vertical magnetic flux increases, the simulation domain becomes more magnetic pressure-dominated and the turbulence develops larger-scale structure. Figure \ref{fig:zpro_beta} shows the corresponding vertical profile of the horizontally- and time-averaged plasma-$\beta$. All of the simulations show the formation of a low-density magnetic pressure-dominated corona  \citep{MillerStone2000}.  The vertical location where the $\langle \widehat{\beta} \rangle_{t} = 1$ transition occurs moves closer to the disc mid-plane with increasing net vertical magnetic flux. The entire domain, including the mid-plane, becomes magnetic pressure-dominated for simulations NVF-$\beta 2$ and NVF-$\beta 1$.

Our two strongest net field runs become fully magnetically dominated, reducing the mid-plane density and pressure in the saturated state as compared to the initial conditions. As a result, while the average vertical magnetic flux is a conserved quantity, the mid-plane value of $\beta$ associated with the vertical field is not. The reduction in $\beta_0^{\rm mid}$ due to the onset of magnetic pressure support means that it is conceivable that an initially unstable disc, with a vertical flux close to the linear stability threshold, could evolve into a magnetically dominated state where the vertical field was no longer unstable. We see no evidence for such an effect, though a simple estimate suggests that our strongest net field simulation does access a regime where the tension of the vertical field is dynamically highly important on all scales. Nominally, magnetic tension from a strong vertical magnetic field will suppress MRI when $v_{{\rm A}, z}^{2} \gtrsim 6 c_{\rm s}^{2} / \pi^{2}$ \citep{BalbusHawley1998}, where $v_{{\rm A}, z}$ is the vertical component of the Alfv\'{e}n speed, $v_{\rm A} = B / \sqrt{\rho}$. In the context of our simulations this stability criterion can be written as $\beta_{z} \lesssim 3.3$. Simulation NVF-$\beta 1$, with $\langle \beta_{z}^{\rm mid} \rangle_{t} = 1.8 \pm 0.3$, is sufficiently magnetized as to approach this limit, but only in the saturated state whose structure is dramatically different from the equilibrium used to compute the linear stability threshold. To quantify this, we measure the variance of the horizontally-averaged gas density at every snapshot in time for each simulation, weighted by mass in order to account for how mass is distributed across regions of a given density,
\begin{equation}
\langle \sigma_{m}^{2} \rangle_{xy} = \left\langle \frac{\rho^{3}}{\langle \rho \rangle_{xy}} \right\rangle_{xy} - \left\langle \frac{\rho^{2}}{\langle \rho \rangle_{xy}} \right\rangle_{xy}^{2}.
\end{equation}
Figure \ref{fig:zpro_densfluc} shows the vertical profile of the horizontally- and time-averaged mass-weighted gas density fluctuations, $\delta \rho_{m} = \sqrt{\sigma_{m}^{2}}$, normalized to the horizontally- and time-averaged mass-weighted gas density vertical profile.  For the zero-net field\footnote{The boundary features seen in Figures \ref{fig:zpro_beta} and \ref{fig:zpro_densfluc} for simulation ZNVF are a consequence of a non-negligible fraction of grid zones reaching the imposed gas density floor.} and weak field runs, the r.m.s. density fluctuations near the mid-plane are at the 10\% level, consistent with the usual expectation that compressibility is ignorable for mid-plane MRI dynamics. (Note that even in these simulations, substantially stronger density fluctuations occur in the corona.) For NVF-$\beta 2$ and NVF-$\beta 1$, conversely, the density fluctuation amplitude is extremely strong, with $\delta \rho / \rho \approx 0.6 - 1$.

To further investigate the inhomogeneous nature of strongly magnetized discs, we compute the normalized probability density function (PDF) of the gas density relative to its mean value at a given height in the disc.  Figure \ref{fig:dens_PDF} shows these PDFs at the disc mid-plane ($z / H_{0} = 0$; {\it left panel}) and in the corona ($z / H_{0} = 2.5$; {\it right panel}).  In other words, Figure \ref{fig:dens_PDF} shows what fraction of volume at a given height is occupied by gas with density $\rho / \langle \rho \rangle_{xy}$.  For the zero-net field and weak field runs, the gas density PDFs are narrowly concentrated about the mean value, which indicates a relatively homogeneous density structure.  However, the moderate and strong field runs have an inhomogeneous density structure, as evidenced by the peak in the gas density PDF shifting to lower $\rho / \langle \rho \rangle_{xy}$ values and the distribution broadening.  Interestingly, a two-component structure develops for the strongest net field simulation NVF-$\beta 1$.  As expected for a disc with $\alpha \sim 1$, NVF-$\beta 1$ has become highly inhomogeneous, such that the relevance of the linear stability analysis for the vertical field is limited.  Indeed, we cannot exclude the possibility that clumpy MRI-unstable discs with even stronger net fields might be possible, at least in situations where the net field evolves slowly toward higher values.

\begin{table}
\addtolength{\tabcolsep}{-4pt}
\centering
\begin{tabular}{c c c c c c c}
\hline
\hline
ID & $\langle \widehat{\beta}^{\rm mid} \rangle_{t}$ & $\langle \widehat{\beta}_{y}^{\rm mid} \rangle_{t}$ & $\langle \widehat{\beta}_{p}^{\rm mid} \rangle_{t}$ & $\langle \widehat{\alpha}^{\rm mid} \rangle_{t}$ & $\langle \widehat{\alpha}_{\rm Rey}^{\rm mid} \rangle_{t}$ & $\langle \widehat{\alpha}_{\rm Max}^{\rm mid} \rangle_{t}$ \\
\hline
ZNVF & $68(6)$ & $81(7)$ & $1.6(1){\rm e}3$ & $0.0088(7)$ & $0.0020(2)$ & $0.0067(5)$ \\
NVF-$\beta 5$ & $41(3)$ & $49(3)$ & $9.3(6){\rm e}2$ & $0.014(1)$ & $0.0034(3)$ & $0.0111(8)$ \\
NVF-$\beta 4$ & $17(1)$ & $21(2)$ & $2.9(4){\rm e}2$ & $0.035(3)$ & $0.008(1)$ & $0.027(2)$ \\
NVF-$\beta 3$ & $2.1(9)$ & $3(1)$ & $3(1){\rm e}1$ & $0.3(1)$ & $0.06(2)$ & $0.22(9)$ \\
NVF-$\beta 2$ & $0.4(1)$ & $0.5(2)$ & $7(2)$ & $1.0(4)$ & $0.16(8)$ & $0.9(3)$ \\
NVF-$\beta 1$ & $0.31(8)$ & $0.5(1)$ & $2.8(5)$ & $1.1(3)$ & $0.12(5)$ & $1.0(3)$ \\
\hline
\end{tabular}
\caption{Fundamental properties of the MRI turbulence, all evaluated at the disc mid-plane and time-averaged.  From {\it left} to {\it right} the columns are: simulation identification label, plasma-$\widehat{\beta}$ parameter, toroidal component of $\widehat{\beta}$, poloidal component of $\widehat{\beta}$, effective $\widehat{\alpha}$-viscosity parameter, Reynolds component of $\widehat{\alpha}$, and Maxwell component of $\widehat{\alpha}$.  Parentheses indicate the $\pm 1 \sigma$ range on the last digit.}
\label{tab:MRI_properties}
\end{table}

\section{Properties of the MRI-Dynamo}
\label{sec:MRIdynamo}
Having established the basic properties of MRI turbulence in our simulations, this section quantifies the properties of the MRI-dynamo.

\subsection{Toroidal Magnetic Field Reversals}
\label{sec:reversals}

\begin{table}
\addtolength{\tabcolsep}{-0pt}
\centering
\begin{tabular}{c c c c c c}
\hline
\hline
ID & $P_{\rm dyn}$ & $\xi$ & $\langle \langle \eta \rangle_{V} \rangle_{t}$ & $\langle \alpha_{B}^{\rm mid} \rangle_{t}$ & $\langle \beta_{\rm B15}^{\rm mid} \rangle_{t}$ \\
& [${\rm orbits}$] & & & & \\
\hline
ZNVF & $12(3)$ & $0.5(1)$ & $0.029(4)$ & $0.016(2)$ & $7(2){\rm e}1$ \\
NVF-$\beta 5$ & $10(2)$ & $0.6(1)$ & $0.043(3)$ & $0.029(3)$ & $4(1){\rm e}1$ \\
NVF-$\beta 4$ & $8(2)$ & $0.7(2)$ & $0.068(6)$ & $0.09(1)$ & $1.7(4){\rm e}1$ \\
NVF-$\beta 3$ & $15(4)$ & $0.4(1)$ & $0.12(2)$ & $0.3(1)$ & $2(1)$ \\
NVF-$\beta 2$ & $8(5){\rm e}1$ & $0.08(5)$ & $0.21(3)$ & $0.3(2)$ & $0.7_{-0.7}^{+1}$ \\
NVF-$\beta 1$ & $-$ & $-$ & $0.30(5)$ & $0.4_{-0.4}^{+0.2}$ & $0.6_{-0.6}^{+1}$ \\
\hline
\end{tabular}
\caption{Fundamental properties of the MRI-dynamo.  From {\it left} to {\it right} the columns are: simulation identification label, period of MRI-dynamo cycles ($P_{\rm dyn}$), $\xi$ parameter (see Equation \ref{eqn:Pdyn}), volume- and time-averaged efficiency parameter for the rise of toroidal magnetic flux ($\eta$), disc mid-plane evaluated and time-averaged parameter for the toroidal magnetic flux production ($\alpha_{B}$), and disc mid-plane evaluated and time-averaged plasma-$\beta$ predicted from Equation \ref{eqn:beta_B15} ($\beta_{\rm B15}^{\rm mid}$).  All of these parameters are defined in the text.  Parentheses indicate the $\pm 1 \sigma$ range on the last digit.}
\label{tab:MRI_dynamo}
\end{table}

\begin{figure*}
  \includegraphics[width=177.8mm]{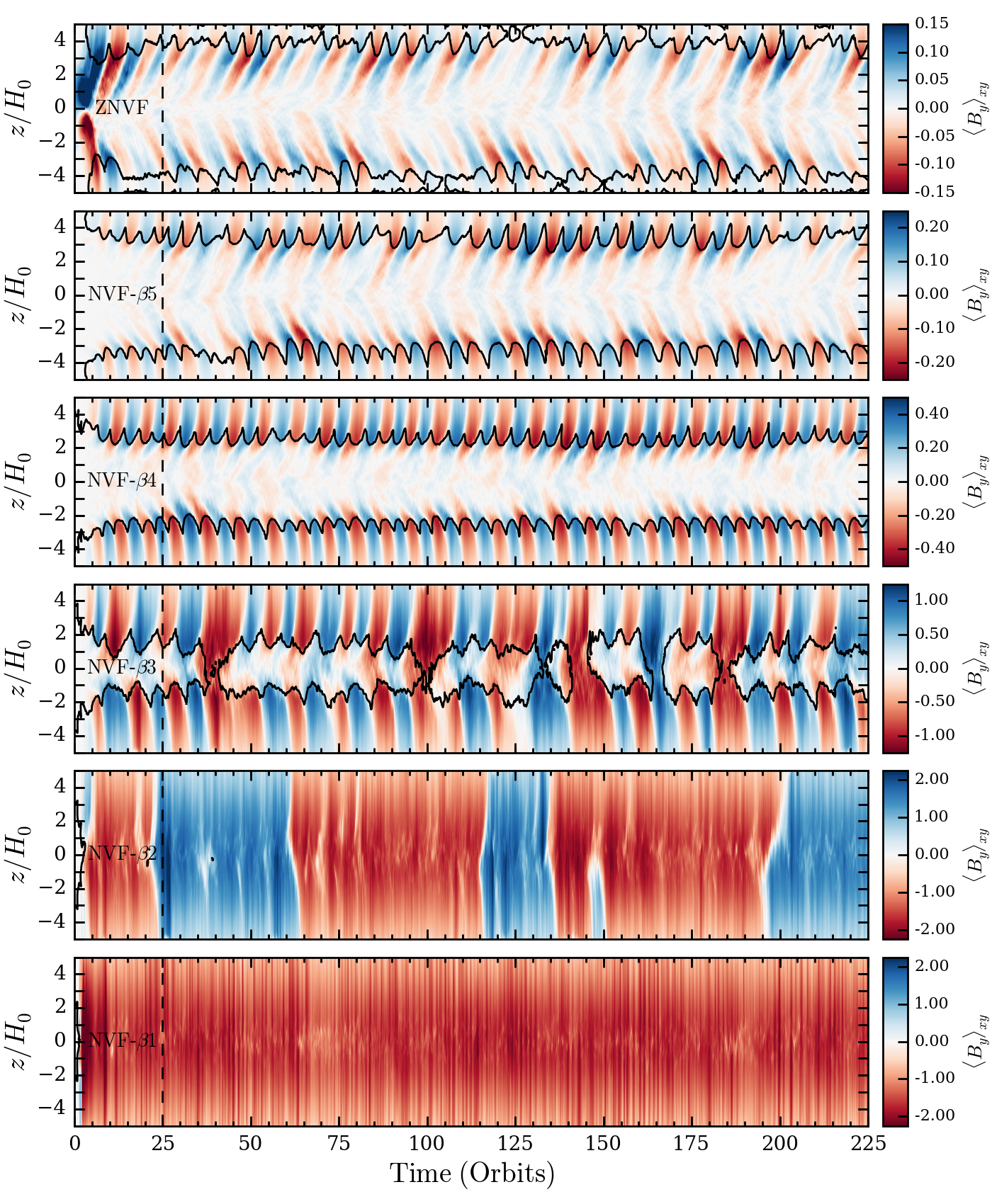}
  \caption{Time evolution of the horizontally-averaged toroidal magnetic field, $\langle B_{y} \rangle_{xy}$, with height from the disc mid-plane, $z / H_{0}$.  From {\it top} to {\it bottom} are space-time diagrams for simulations ZNVF, NVF-$\beta 5$, NVF-$\beta 4$, NVF-$\beta 3$, NVF-$\beta 2$, NVF-$\beta 1$.  MRI-dynamo behaviour occurs in all but the most strongly magnetized simulation NVF-$\beta 1$.  The period of the MRI-dynamo cycle generally becomes longer and reversals become more sporadic with increasing net vertical magnetic flux.  {\it Black lines} show the $\widehat{\beta} = 1$ contour.  The entire disc becomes magnetic pressure-dominated for brief episodes in simulation NVF-$\beta 3$ and for all times in simulations NVF-$\beta 2$ and NVF-$\beta 1$.}
  \label{fig:butterfly}
\end{figure*}

\begin{figure*}
  \includegraphics[width=84mm]{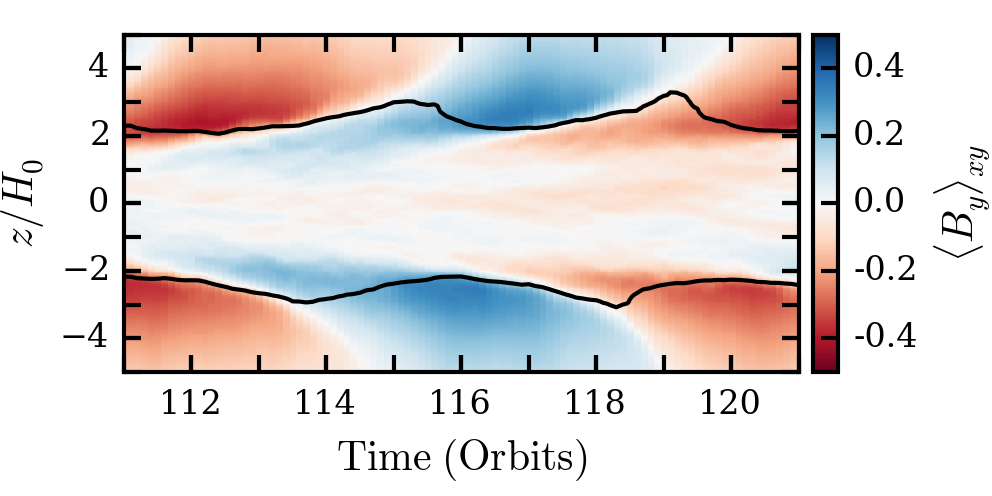}  
  \includegraphics[width=84mm]{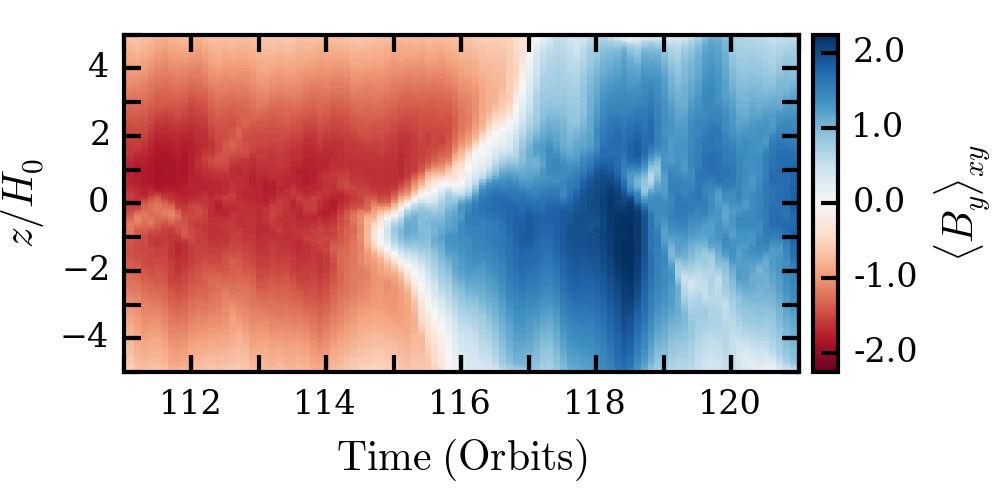}  
  \caption{Butterfly diagrams for simulations NVF-$\beta 4$ ({\it left}) and NVF-$\beta 2$ ({\it right}), zoomed in over 10 orbits to highlight the $\langle B_{y} \rangle_{xy}$ structure during a reversal.  {\it Black lines} show the $\widehat{\beta} = 1$ contour.  The current sheet sweeping through the vertical domain during a reversal in $\langle B_{y} \rangle_{xy}$ propagates faster in strongly magnetized discs ({\it right}) than in weakly magnetized discs ({\it left}).}
  \label{fig:butterfly_zoom}
\end{figure*}

\begin{figure*}
  \includegraphics[width=42mm]{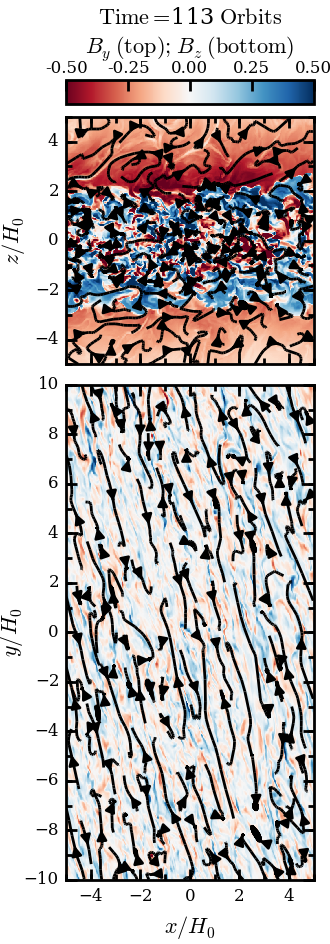}  
  \includegraphics[width=42mm]{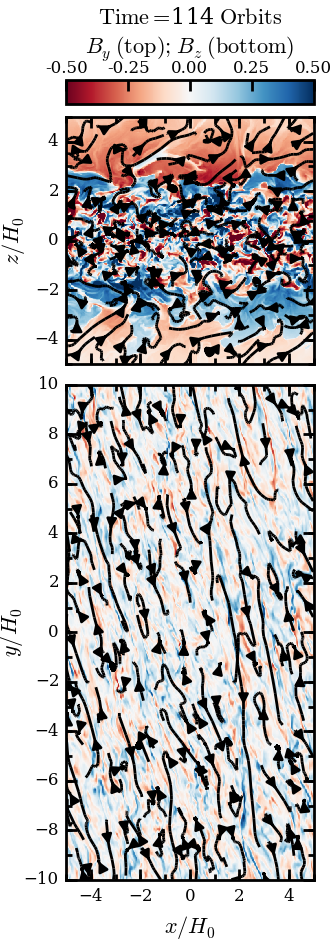}  
  \includegraphics[width=42mm]{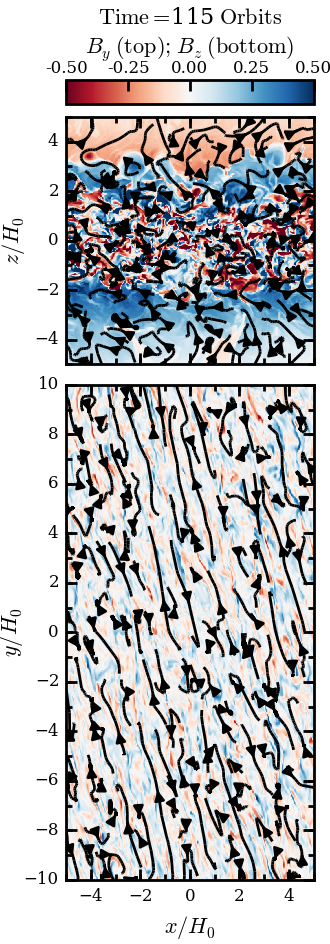}  
  \includegraphics[width=42mm]{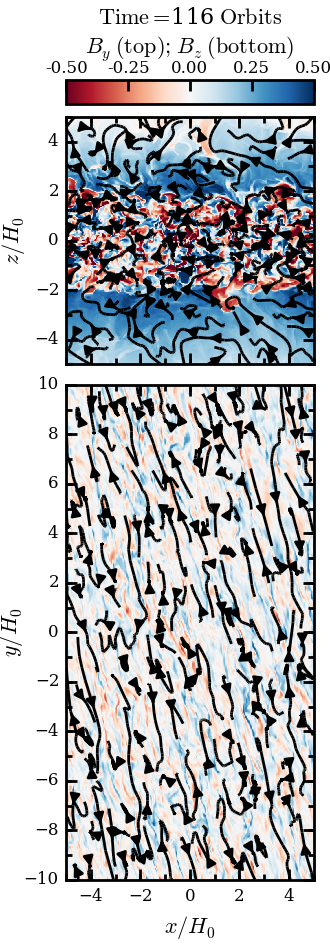}  
  \caption{Magnetic field structure during an MRI-dynamo reversal in the large-scale toroidal magnetic field (see {\it left panel} of Figure \ref{fig:butterfly_zoom}) for the weak net vertical magnetic flux simulation NVF-$\beta 4$.  The {\it top panels} show colour renderings of the toroidal magnetic field and poloidal magnetic field lines for a slice through the $xz$-plane at $y = 0$.  The {\it bottom panels} show colour renderings of the vertical magnetic field and horizontal magnetic field lines for a slice through the $xy$-plane at $z = 0$.  Panels from {\it left} to {\it right} show times $t = 113, 114, 115, 116$ orbits.  $B_{y}$ is fairly turbulent, but relatively well-organized, near the equator and is more organized above/below the disc mid-plane.}
  \label{fig:Blines_Bweak}
\end{figure*}

\begin{figure*}
  \includegraphics[width=42mm]{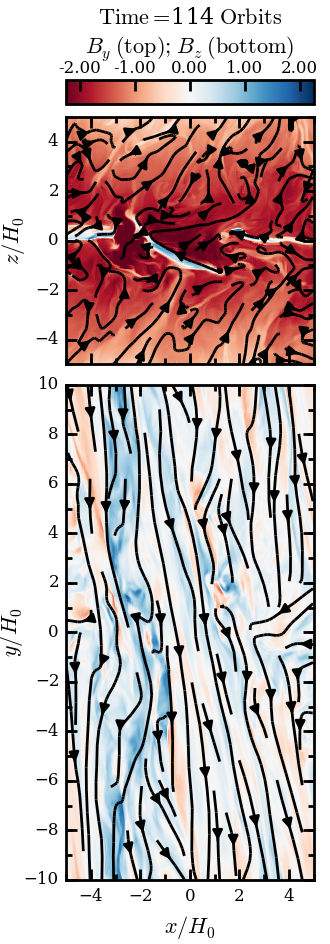}  
  \includegraphics[width=42mm]{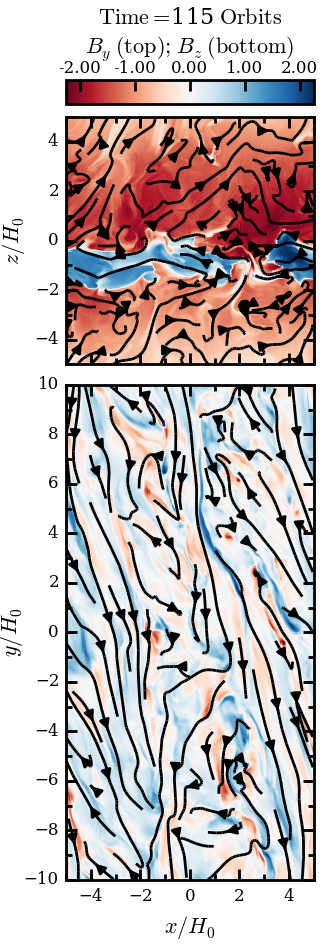}  
  \includegraphics[width=42mm]{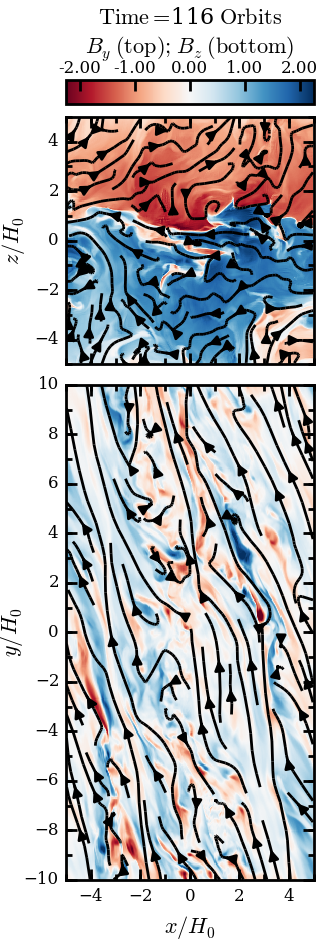}  
  \includegraphics[width=42mm]{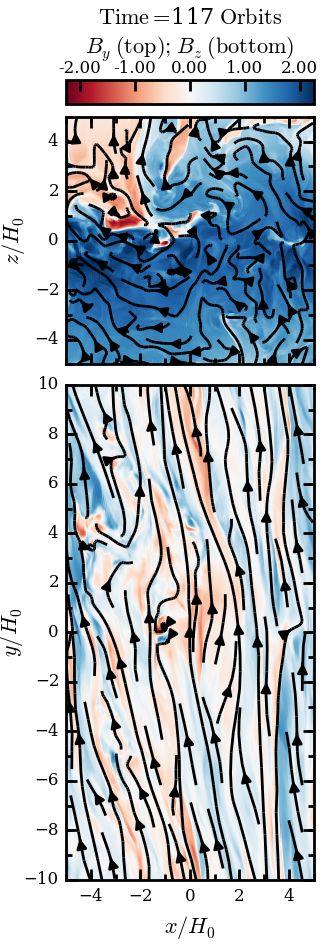}  
  \caption{Magnetic field structure during an MRI-dynamo reversal in the large-scale toroidal magnetic field (see {\it right panel} of Figure \ref{fig:butterfly_zoom}) for the strong net vertical magnetic flux simulation NVF-$\beta 2$.  The {\it top panels} show colour renderings of the toroidal magnetic field and poloidal magnetic field lines for a slice through the $xz$-plane at $y = 0$.  The {\it bottom panels} show colour renderings of the vertical magnetic field and horizontal magnetic field lines for a slice through the $xy$-plane at $z = 0$.  Panels from {\it left} to {\it right} show  times $t = 114, 115, 116, 117$ orbits.  Organized $B_{y}$ structures dominate throughout the domain and are highly organized on the equator.}
  \label{fig:Blines_Bstrong}
\end{figure*}

\begin{figure*}
  \includegraphics[width=84mm]{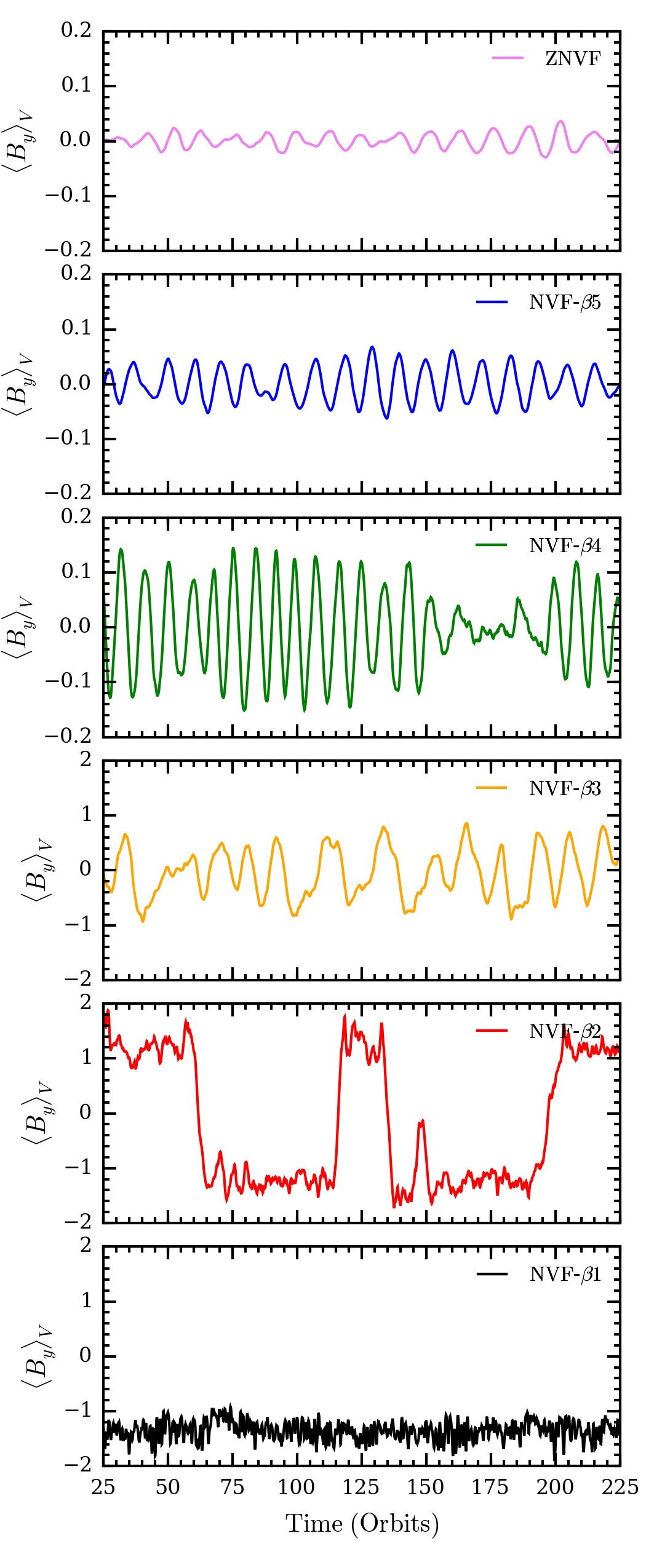}
  \includegraphics[width=84mm]{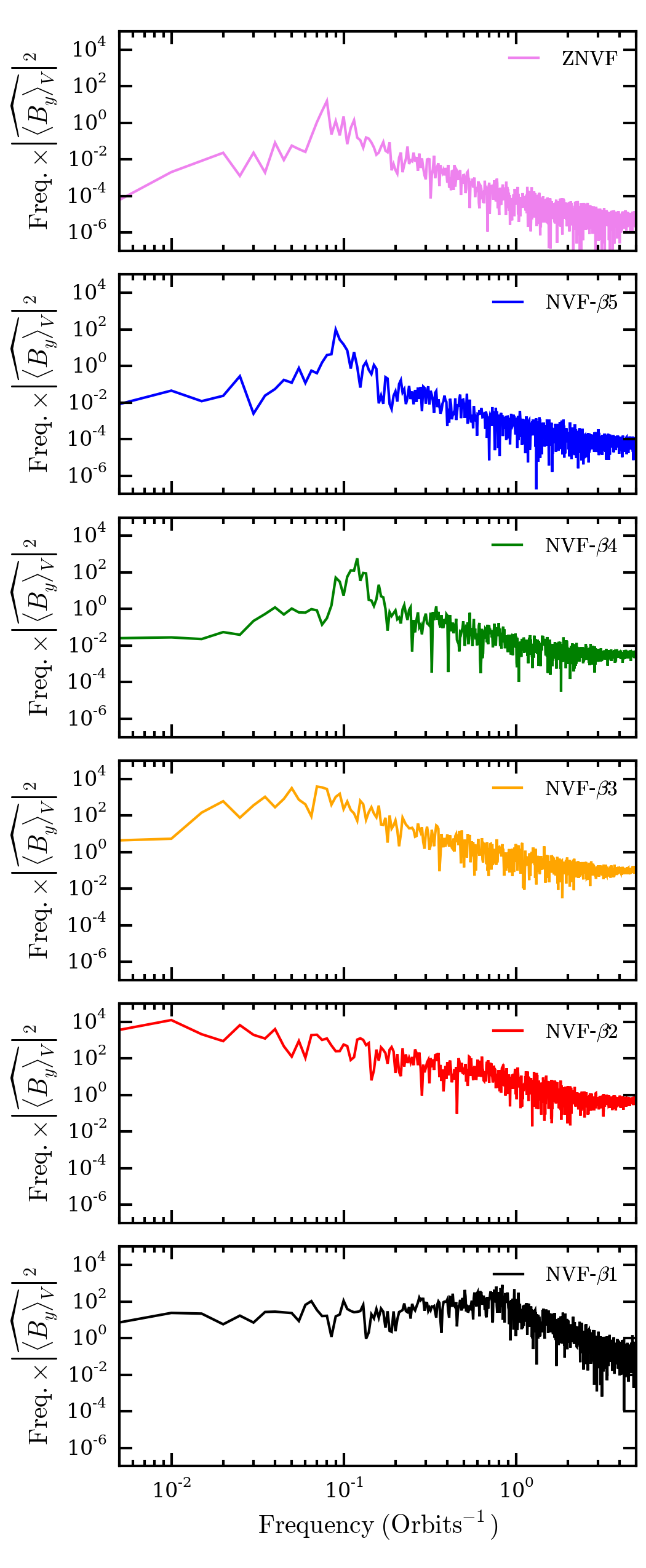}
  \caption{Time evolution of the volume-averaged toroidal magnetic field ({\it left panels}) and its power spectrum ({\it right panels}).  From {\it top} to {\it bottom} are simulations ZNVF, NVF-$\beta 5$, NVF-$\beta 4$, NVF-$\beta 3$, NVF-$\beta 2$, NVF-$\beta 1$.  For simulations with a zero or weak ($\beta_{0}^{\rm mid} \lesssim 10^{4}$) net vertical magnetic flux, a coherent peak appears in the power spectrum resulting from a regular MRI-dynamo period of $P_{\rm dyn} \simeq 10$ orbits.  This feature is substantially broadened for the moderately magnetized simulation NVF-$\beta 3$ due to the MRI-dynamo flips becoming more sporadic and having a longer period.  The strongly magnetized simulation NVF-$\beta 2$ shows no discernible power spectral peak.  The very strongly magnetized simulation NVF-$\beta 1$ does not exhibit MRI-dynamo cycles and has a flat power spectrum with a high-frequency cutoff at $\simeq 1~{\rm orbit^{-1}}$.}
  \label{fig:powspec_By}
\end{figure*}

For each simulation, Figure \ref{fig:butterfly} shows the space-time diagram of the horizontally-averaged toroidal magnetic field, $\langle B_{y} \rangle_{xy}$.  These so-called ``butterfly diagrams'' exhibit quasi-periodic reversals of $\langle B_{y} \rangle_{xy}$, which is a hallmark of accretion disc MRI-dynamo activity \citep[e.g.,][]{Brandenburg1995}.  The amount of net vertical magnetic flux increases from {\it top} to {\it bottom} in Figure \ref{fig:butterfly} and {\it black contours} show the transitional vertical location where $\widehat{\beta} = 1$.  Reversals in the large-scale toroidal magnetic field showcase regular periodicity for the zero net vertical magnetic flux simulation and simulations with $\beta_{0}^{\rm mid} \gtrsim 10^{4}$.  This is a well-known result for stratified shearing boxes \citep[e.g.,][]{Simon2012} and global disc simulations \citep[e.g.,][]{ONeill2011} with zero net vertical magnetic flux.  However, previous shearing box simulations with a net vertical magnetic flux and $\beta_{0}^{\rm mid} \gtrsim 10^{4}$ \citep{Fromang2013, BaiStone2013} show highly irregular dynamo patterns.

In the moderately magnetized regime with $\beta_{0}^{\rm mid} = 10^{3}$, Figure \ref{fig:butterfly} shows that the toroidal magnetic field reversals become less frequent and more sporadic, with temporary episodes where the entire disc domain becomes magnetic pressure-dominated.  This behaviour was also observed in the $\beta_{0}^{\rm mid} = 10^{3}$ shearing box simulations of \citet{BaiStone2013}.  In the $\beta_{0}^{\rm mid} = 10^{2}$ strong net vertical magnetic flux simulation, the entire disc domain achieves $\widehat{\beta} < 1$ for all times and the toroidal field reversals become still less frequent and more sporadic. However, they still occur. To our knowledge this is the first time that the MRI-dynamo butterfly pattern has been observed to persist in a magnetic pressure-dominated disc.  The toroidal magnetic field flip-flops disappear for the very strongly magnetized $\beta_{0}^{\rm mid} = 10^{1}$ simulation, though it is possible that the simulation was not evolved to an adequately long time and/or a larger domain size is necessary to observe this behaviour.

Figure \ref{fig:butterfly} also shows that the MRI-dynamo generates an ordered toroidal magnetic field near the disc mid-plane, which buoyantly rises and reaches a maximal amplification near the height where $\widehat{\beta} \simeq 1$.  The colour bars in Figure \ref{fig:butterfly} show that the peak strength that the toroidal magnetic field reaches is similar for the zero net vertical magnetic flux simulation ZNVF and the very weak net vertical magnetic flux simulation NVF-$\beta 5$.  This peak toroidal field strength roughly doubles with each order of magnitude decrease in $\beta_{0}^{\rm mid}$ until the $\beta_{0}^{\rm mid} = 10^{1}$ simulation, where the toroidal field strength appears to saturate to the same level as in the $\beta_{0}^{\rm mid} = 10^{2}$ simulation.  This is suggestive of a threshold value to the toroidal magnetic field that the MRI-dynamo can produce from an imposed net vertical magnetic flux.

The period of MRI-dynamo cycles can be measured as $P_{\rm dyn} = \Delta t / N_{\rm cycles}$.  Over the time interval $\Delta t = t_{\rm f} - t_{\rm i} = 200$ orbits, we measure the number of MRI-dynamo cycles, $N_{\rm cycles} = N_{\rm flips} / 2$, by counting the number of $\langle B_{y} \rangle_{V}$ reversals separately in the top ($z > 0$) and bottom ($z < 0$) simulation domains and averaging them together to get $N_{\rm flips} = \left( N_{\rm flips}^{\rm top} + N_{\rm flips}^{\rm bot} \right) / 2$.  We measure $P_{\rm dyn}$ and estimate its uncertainty as $P_{\rm dyn} / \sqrt{N_{\rm cycles}}$.  The MRI-dynamo period (in units of orbits) can also be parametrized as,
\begin{equation}
P_{\rm dyn} = \frac{2 \pi}{\xi \Omega}~\left[ {\rm orbits} \right], \label{eqn:Pdyn}
\end{equation}
and we list the results for $P_{\rm dyn}$ and $\xi$ in Table \ref{tab:MRI_dynamo}.  The $\xi$ parameter will be important in \S \ref{sec:magflux} and \S \ref{sec:diskstruct}.

Figure \ref{fig:butterfly_zoom} shows zoomed-in subsets of Figure \ref{fig:butterfly} for simulations NVF-$\beta 4$ ({\it left}) and NVF-$\beta 2$ ({\it right}) to highlight the space-time structure of toroidal magnetic field reversals for weak and strong net vertical magnetic flux cases, respectively.  Notably, the rise speed of the current sheets launched during each reversal of the toroidal magnetic field increases with net vertical magnetic flux.

Figures \ref{fig:Blines_Bweak} and \ref{fig:Blines_Bstrong} show the detailed magnetic field structure for simulations NVF-$\beta 4$ and NVF-$\beta 2$, respectively, over the course of the $\langle B_{y} \rangle_{xy}$ reversals highlighted in Figure \ref{fig:butterfly_zoom}.  Comparing Figures \ref{fig:Blines_Bweak} and \ref{fig:Blines_Bstrong} demonstrates that the MRI-dynamo behaves differently in the weak and strong net vertical flux cases.

For the weakly magnetized NVF-$\beta 4$, the azimuthal slice ($xz$-plane) of Figure \ref{fig:Blines_Bweak} shows small-scale, turbulent magnetic field structure concentrated to the disc mid-plane regions, with $B_{y}$ becoming more organized at heights where $\beta \lesssim 1$.  While there is a mildly dominant $B_{y}$ polarity in the mid-plane regions at any given time, a blend of both positive and negative $B_{y}$ structures is always present. The equatorial slice ($xy$-plane) shows that even at the equator, $B_{y}$ is relatively well-organized.

For the strongly magnetized NVF-$\beta 2$, Figure \ref{fig:Blines_Bstrong} shows that $B_{y}$ dominates at all heights and is highly organized at the disc mid-plane.  In the {\it leftmost panel}, a precursor of the MRI-dynamo flip is present in the form of a few ribbon-like positive ({\it blue}) $B_{y}$ structures.  While the entire domain is dominated by either positive or negative $B_{y}$ at a given time preceding or following an MRI-dynamo reversal, wispy $B_{y}$ structures of the non-dominant polarity always persist in the vicinity of the disc mid-plane.  Indeed, the {\it rightmost panel shows that a negative ({\it red}) $B_{y}$ ribbon-like precursor is now present near the mid-plane following an MRI-dynamo flip.} For both NVF-$\beta 4$ and NVF-$\beta 2$, the MRI-dynamo launches current sheets into the corona --- although they do not propagate perfectly synchronously above/below the disc mid-plane --- and the large-scale toroidal magnetic field in the coronal region reverses sign.

Several authors have attempted to understand the origin of the MRI-dynamo field reversals, and quantitative properties such as the period, using $\alpha \Omega$ dynamo theory \citep[see e.g.][and references therein]{GresselPessah2015}. In a different approach, \cite{Herault2011} computed exactly time-periodic dynamo solutions to an incompressible model at modest Reynolds and magnetic Reynolds numbers. They argued that the MRI-dynamo results from interactions between the dominant toroidal field and non-axisymmetric perturbations. The relevance of these simplified models to our simulations --- which as we have noted are highly compressible when the net flux is strong --- is unclear, but it is possible that sufficient small-scale turbulence is a prerequisite for MRI-dynamo reversals.  We observe a clear decrease in small-scale turbulence as the net field increases (see Figures \ref{fig:Blines_Bweak} and \ref{fig:Blines_Bstrong}).

The {\it left panel} of Figure \ref{fig:powspec_By} shows the time-evolution of the volume-averaged toroidal magnetic field, $\langle B_{y} \rangle_{V}$, for each simulation.  The MRI-dynamo cycles are apparent from the sinusoidal behaviour of $\langle B_{y} \rangle_{V}$.  A dynamo flip in one disc hemisphere is nearly always mirrored by a similar response in the other disc hemisphere; however, this response is not necessarily instantaneous.  For instance, consider the time range $t \simeq 150 - 200$ in simulation NVF-$\beta 4$.  Figure \ref{fig:butterfly} shows a well-defined butterfly pattern for $\langle B_{y} \rangle_{xy}$, but Figure \ref{fig:powspec_By} shows that the top ($z > 0$) and bottom ($z < 0$) domains are anti-synchronized, as evidenced by $\langle B_{y} \rangle_{V} \simeq 0$ in this time range.

The {\it right panel} of Figure \ref{fig:powspec_By} shows the temporal power spectrum generated from $\langle B_{y} \rangle_{V}$ for each simulation.  These are not significantly altered by the occasional offset nature of $\langle B_{y} \rangle_{V}$ in the top and bottom disc hemispheres just described.  The weakly magnetized simulations with a well-defined MRI-dynamo pattern (ZNVF, NVF-$\beta 5$, NVF-$\beta 4$) display a coherent peak in the power spectrum, while this feature broadens/disappears for the strongly magnetized simulations because the MRI-dynamo cycles are either sporadic (NVF-$\beta 3$, NVF-$\beta 2$) or non-existent (NVF-$\beta 1$).  

\subsection{Production and Escape of Toroidal Magnetic Field}
\label{sec:magflux}

\begin{figure}
  \includegraphics[width=84mm]{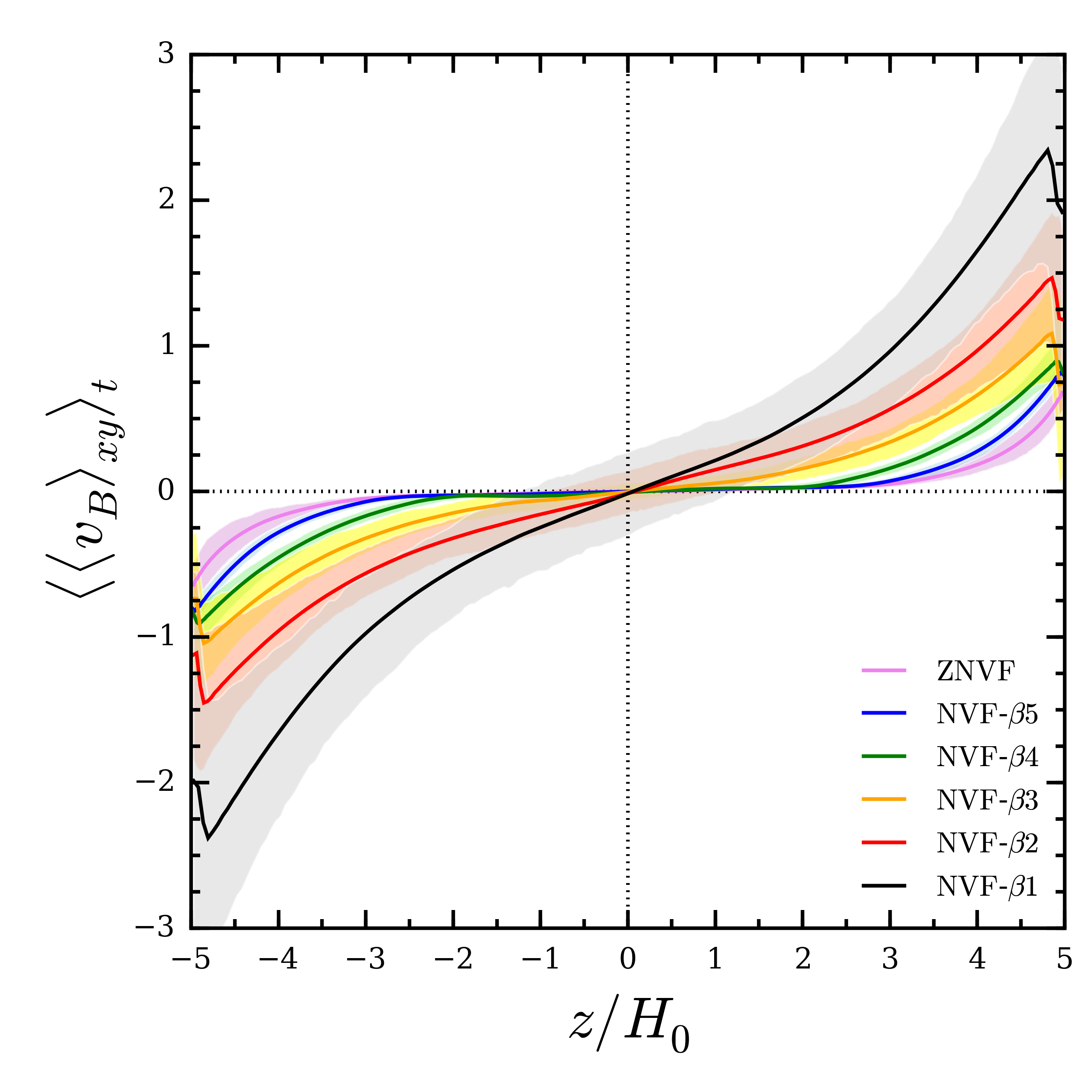}
  \caption{Vertical profiles of the horizontally- and time-averaged rise speed (in code units) of the toroidal magnetic flux (see Figure \ref{fig:alpha_tevol} for line colour conventions).  For all simulations, $v_{B} \simeq 0$ within the gas-pressure dominated regions (i.e., interior to the $z$-locations where $\langle \widehat{\beta} \rangle_{t} \simeq 1$).  Outside of this vertical location, $v_{B} > 0$ and increases with disc altitude.  As the net vertical magnetic flux increases, $v_{B}$ increases and even reaches supersonic speeds (i.e., $v_{B} > c_{\rm s} = 1$) for simulations NVF-$\beta 2$ and NVF-$\beta 1$.}
  \label{fig:zpro_vB}
\end{figure}

The MRI-dynamo channels accretion power into ordered toroidal magnetic fields near the disc mid-plane, which buoyantly rise through the disc atmosphere and escape.  In the model for the MRI-dynamo proposed by \citet{Begelman2015}, accretion energy is being liberated locally by the MRI and some fraction of this is channeled into the toroidal magnetic field.  This toroidal field rises due to the Parker instability and carries a vertical flux of magnetic energy (i.e., Poynting flux).  However, this buoyant rise of toroidal field is hampered by field buckling and reconnection, causing energy to be lost to the gas by the magnetic field as it rises.  The details of this complicated energy exchange process are simplified by assuming that all of the energy lost by the toroidal magnetic field as it rises goes into locally heating the gas.  In this model for toroidal magnetic flux production and escape, the governing energy balance equation is,
\begin{equation}
\frac{d}{d z} \left( p_{B} v_{B}^{2} \right) = \Omega v_{B} \left[ \alpha_{B} \left( p_{\rm gas} + p_{B} \right) - 2 \xi p_{B} \right], \label{eqn:energy_balance}
\end{equation}
where $\alpha_{B}$ measures the portion of accretion energy liberated locally by the MRI that goes into the toroidal magnetic field and we recall the MRI-dynamo period parameter $\xi$ from Equation \ref{eqn:Pdyn}.  The vertical profile for the rise speed of the toroidal magnetic flux can be expressed as,
\begin{equation}
v_{B} \left( z \right) = \eta \Omega z, \label{eqn:vB_analytic}
\end{equation}
where $\eta = 1$ would correspond to the field rising at the free-fall speed. Values of $\eta \le 1$ phenomenologically account for effects that slow down the rise speed of the toroidal magnetic field, such as tangling of magnetic field lines and magnetic reconnection.

Here, we use our simulations to measure quantities that characterize toroidal magnetic flux production ($\alpha_{B}$) and escape ($v_{\rm B}$, $\eta$) within the context of this model. We measure the horizontally-averaged rise speed of the toroidal magnetic field, $\langle v_{B} \rangle_{xy}$, as follows.  The flux of magnetic energy --- or Poynting flux, $\mathbf{F}_{B} = \mathbf{B} \times \left( \mathbf{v} \times \mathbf{B} \right)$ --- has vertical component,
\begin{equation}
F_{B, z} = B_{x} \left( v_{z} B_{x} - v_{x} B_{z} \right) - B_{y} \left( v_{y} B_{z} - v_{z} B_{y} \right). \label{eqn:FB_z}
\end{equation}
If we assume that the magnetic field is purely toroidal, $B \simeq B_{y}$, which is justified by our simulations, then the $z$-component of the Poynting flux becomes,
\begin{equation}
F_{B, z} = v_{B} B_{y}^{2}, \label{eqn:FB_z_assume}
\end{equation}
where $v_{B} = v_{z}$ is a measure of the rise speed of the toroidal magnetic field defined by Equation \ref{eqn:vB_analytic}.  To measure $v_{B}$ from our simulations, we equate Equations \ref{eqn:FB_z} and \ref{eqn:FB_z_assume}, solve for $v_{B}$, and horizontally-average the result to obtain,
\begin{equation}
\langle v_{B} \rangle_{xy} = \frac{\left\langle B_{x} \left( v_{z} B_{x} - v_{x} B_{z} \right) - B_{y} \left( v_{y} B_{z} - v_{z} B_{y} \right) \right\rangle_{xy}}{\left\langle B_{y}^{2} \right\rangle_{xy}}. \label{eqn:vB}
\end{equation}
Figure \ref{fig:zpro_vB} shows the horizontally- and time-averaged $v_{B}$ for each simulation.  The toroidal magnetic field rise speed is essentially zero in the gas pressure-dominated regions where the disc exhibits small-scale turbulence and the large-scale toroidal magnetic field is not strongly ordered.  For vertical locations exterior to the $\langle \widehat{\beta} \rangle_{t} \simeq 1$ transition (see Figure \ref{fig:zpro_beta}), where the toroidal magnetic field dominates the vertical pressure support in the disc, $v_{B}$ is increasing with disc altitude.  As the amount of net vertical magnetic flux increases, $v_{B}$ at any given height above/below the disc mid-plane also increases, even becoming supersonic for simulations NVF-$\beta 2$ and NVF-$\beta 1$.  Recall that $c_{\rm s} =1$ throughout the domain for our isothermal simulations.

We measure the vertical profile of the toroidal magnetic flux escape speed parameter $\eta \left( z \right)$ by inserting $\langle v_{B} \rangle_{xy}$ from Equation \ref{eqn:vB} into Equation \ref{eqn:vB_analytic}.  We average over $z$ to obtain $\langle \eta \rangle_{V}$, ignoring numerical artifacts concentrated only very close to the disc mid-plane and the vertical boundaries.  Table \ref{tab:MRI_dynamo} lists $\langle \langle \eta \rangle_{V} \rangle_{t}$ for each simulation.  As the amount of net vertical magnetic flux increases, the toroidal magnetic field is able to rise more efficiently (i.e., $\eta$ increases), with the speed reaching 20-30\% of the free-fall speed for the two magnetically dominated runs.  Presumably, this is a consequence of the toroidal magnetic field becoming more ordered with increasing net vertical magnetic flux (see Figures \ref{fig:Blines_Bweak} and \ref{fig:Blines_Bstrong}), i.e., there is less tangling and reconnection of magnetic field lines. 

We measure the vertical profile of the toroidal magnetic flux production parameter $\alpha_{B}$ by rearranging Equation \ref{eqn:energy_balance} to obtain,
\begin{align}
\alpha_{B} \left( z \right) &= \frac{1}{\langle p_{\rm gas} \rangle_{xy} + \langle p_{B} \rangle_{xy}} \nonumber \\
&\times \left[ \frac{1}{\Omega} \left( \langle v_{B} \rangle_{xy} \frac{d \langle p_{B} \rangle_{xy}}{d z} + 2 \langle p_{B} \rangle_{xy} \frac{d \langle v_{B} \rangle_{xy}}{d z} \right) + 2 \xi \langle p_{B} \rangle_{xy} \right]. \label{eqn:alphaB_analytic}
\end{align}
In the vertical locations near the disc mid-plane $\alpha_{B}$ has an essentially flat profile, which is not shown here for the sake of brevity.  We adopted a constant value for $\xi$; therefore our measured values for $\alpha_{B}$ account for energy dissipation during MRI-dynamo flips in a volume-averaged sense and not as a function of disc altitude.  Given that $\alpha_{B}$ parametrizes the amount of accretion power liberated by the MRI that goes into the toroidal magnetic field and that we measure $\alpha_{B}$ to have a flat vertical profile in the MRI-active disc mid-plane regions (i.e., $| z / H_{0} | < 1$), we choose to evaluate $\alpha_{B}$ at the disc mid-plane. Table \ref{tab:MRI_dynamo} lists $\langle \alpha_{B}^{\rm mid} \rangle_{t}$ for each simulation.  As the amount of the net vertical magnetic flux increases, the MRI channels a fractionally larger amount of the liberated accretion energy into toroidal magnetic fields.

Notably, as shown by inspecting Equation \ref{eqn:alphaB_analytic}, the values reported for $\alpha_{B}$ are heavily influenced by the frequency of MRI-dynamo cycles, as parametrized by $\xi$.  A current sheet sweeps vertically through the domain with each MRI-dynamo reversal in the large-scale toroidal magnetic field.  These strong current sheets are likely sites of magnetic reconnection and could potentially deposit substantial amounts of accretion power in the coronal regions, presumably through heating by magnetic reconnection.  For the zero and weak/moderate ($\beta_{0}^{\rm mid} \gtrsim 10^{3}$) net vertical magnetic flux simulations, this heating caused by MRI-dynamo cycles dominates the $\alpha_{B}$ values we measure.  For the NVF-$\beta 2$ simulation, MRI-dynamo cycles are infrequent and contribute in an ancillary way to $\alpha_{B}$.

By considering the interplay of toroidal magnetic flux production and its buoyant escape, along with the effects of heating by current sheets launched during each MRI-dynamo cycle, \citet{Begelman2015} show that the plasma-$\beta$ at the disc mid-plane can be determined via a regularity condition,
\begin{equation}
\beta_{\rm B15}^{\rm mid} = \frac{2 \eta}{\alpha_{B}} + \nu -1, \label{eqn:beta_B15}
\end{equation}
where $\nu \equiv 2 \xi / \alpha_{B}$ is the reconnection efficiency parameter.  For each simulation, Table \ref{tab:MRI_dynamo} gives the values of $\langle \beta_{\rm B15}^{\rm mid} \rangle_{t}$ calculated  from the values of $\xi$, $\eta$, and $\alpha_{B}$ also listed in Table \ref{tab:MRI_dynamo}.  Comparing these model-dependent values for the mid-plane plasma-$\beta$ to those that we measure {\it directly} in our simulations (see $\langle \widehat{\beta}^{\rm mid} \rangle_{t}$ in Table \ref{tab:MRI_properties}), we find extremely good agreement. The mid-plane magnetization can thus be understood in terms of a very simple model that balances toroidal field production against vertical escape at a rate tied to the free-fall speed, though simulations are needed (especially for weak net fields) to measure the efficiency of these processes as a function of the net field strength.

\subsection{Vertical Disc Structure}
\label{sec:diskstruct}

\begin{figure*}
  \includegraphics[width=84mm]{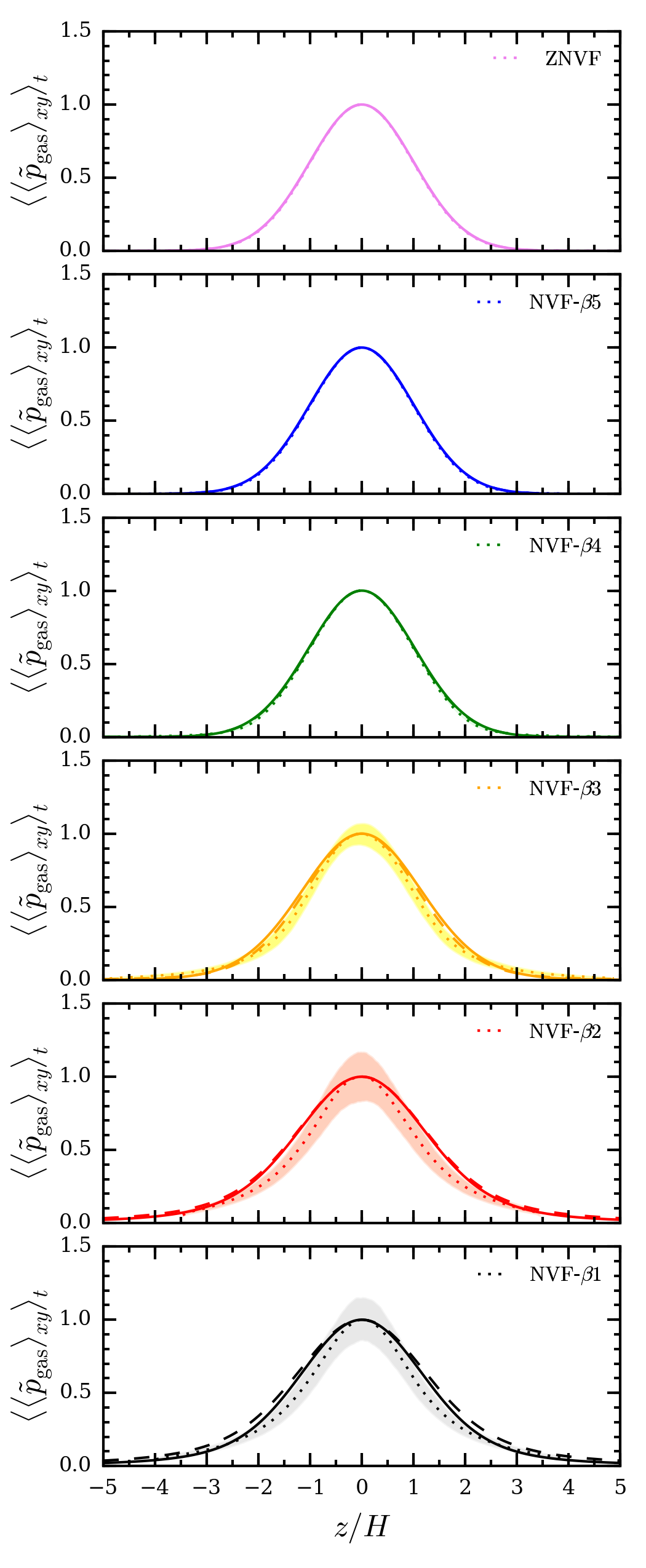}
  \includegraphics[width=84mm]{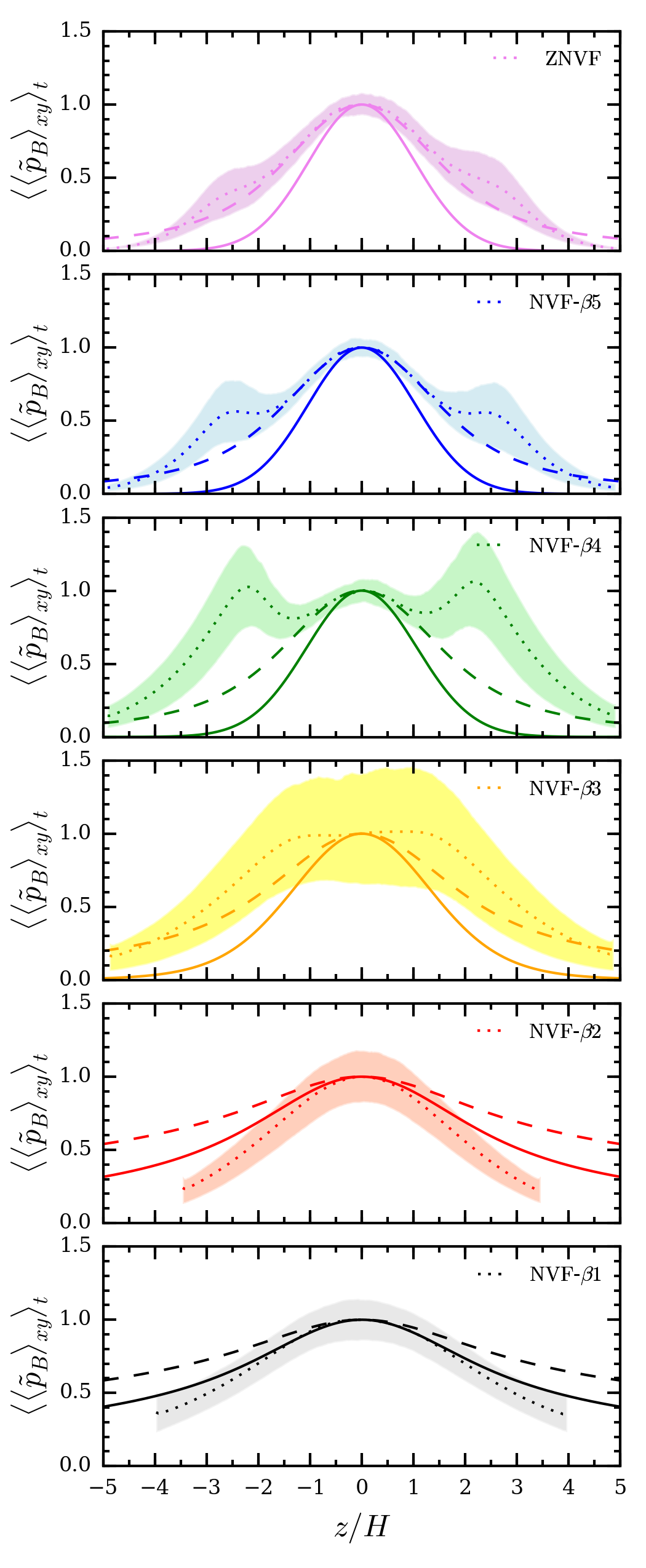}
  \caption{{\it Dotted lines} show the vertical profiles of the horizontally- and time-averaged gas pressure ({\it left panel}) and magnetic pressure ({\it right panel}), each normalized to their value at the disc mid-plane (see Figure \ref{fig:alpha_tevol} for line colour conventions).  Also plotted are the analytic model predictions \citep[Equations \ref{eqn:hydrostat} and \ref{eqn:Poynting};][]{Begelman2015}: {\it solid lines} adopt $\beta^{\rm mid} = \langle \beta_{\rm B15}^{\rm mid} \rangle_{t}$ and $\nu = 2 \xi / \langle \alpha_{B}^{\rm mid} \rangle_{t}$ from Table \ref{tab:MRI_dynamo}; {\it dashed lines} adopt $\beta^{\rm mid} = \langle \widehat{\beta}^{\rm mid} \rangle_{t}$ from Table \ref{tab:MRI_properties} and $\nu = 0$.  From {\it top} to {\it bottom} are simulations ZNVF, NVF-$\beta 5$, NVF-$\beta 4$, NVF-$\beta 3$, NVF-$\beta 2$, NVF-$\beta 1$.  We note that $z$ is plotted in units of the gas density scale height, $H$, that develops for each individual simulation and not its initial value, $H_{0}$.}
  \label{fig:zpro_p_pB}
\end{figure*}

We can also compare our simulation results to the analytic predictions for the vertical disc structure in the context of the MRI-dynamo model discussed in \S \ref{sec:magflux}. Figure \ref{fig:zpro_p_pB} shows the vertical profiles of the horizontally- and time-averaged gas pressure ({\it left panels}) and magnetic pressure ({\it right panels}), where disc altitude is in units of the gas density scale height $H$ that develops in the turbulent steady state of each simulation and not its initial value $H_{0}$.  Given the isothermal equation of state that we adopt, the gas pressure profiles are equivalent to the gas density profiles.

Using Equation \ref{eqn:beta_B15} and combining equations for hydrostatic equilibrium and the Poynting flux, \citet{Begelman2015} derive coupled ordinary differential equations that govern the vertical structure,
\begin{align}
\frac{d}{d y} \left( \tilde{p}_{\rm B} + \beta^{\rm mid} \tilde{p}_{\rm gas} \right) &= - \beta^{\rm mid} \tilde{\rho} \label{eqn:hydrostat} \\
y \frac{d \tilde{p}_{\rm B}}{d y} &= \frac{\beta^{\rm mid}}{1 + \beta^{\rm mid} - \nu} \left( \tilde{p}_{\rm gas} - \tilde{p}_{\rm B} \right), \label{eqn:Poynting}
\end{align}
where $\tilde{\rho} = \rho / \rho^{\rm mid}$, $\tilde{p}_{\rm gas} = p_{\rm gas} / p_{\rm gas}^{\rm mid}$, $\tilde{p}_{B} = p_{B} / p_{B}^{\rm mid}$, and $y = z^{2} / (2 H^{2})$.

Combining Equations \ref{eqn:hydrostat} and \ref{eqn:Poynting} with an isothermal equation of state, Figure \ref{fig:zpro_p_pB} shows the vertical profiles of gas pressure ({\it left panels}) and magnetic pressure ({\it right panels}) predicted by the \citet{Begelman2015} model and compared to the simulation results.  The {\it solid lines} show the results from choosing $\beta^{\rm mid} = \langle \beta_{\rm B15}^{\rm mid} \rangle_{t}$ and $\nu = 2 \xi / \langle \alpha_{B}^{\rm mid} \rangle_{t}$ (see Table \ref{tab:MRI_dynamo}).  The {\it dashed lines} show the results from choosing $\beta^{\rm mid} = \langle \widehat{\beta}^{\rm mid} \rangle_{t}$ (see Table \ref{tab:MRI_properties}) and $\nu = 0$.  The main effect of including the magnetic reconnection efficiency parameter $\nu$ is to steepen the magnetic pressure profile.  The model that does not include $\nu$ ({\it dashed lines}) roughly matches the magnetic pressure profiles for the zero and weak/moderate ($\beta_{0}^{\rm mid} \gtrsim 10^{3}$) net vertical magnetic flux simulations.  The model that does include $\nu$ ({\it solid lines}) does a better job of capturing the steepening of the magnetic pressure profiles observed in the magnetic pressure-dominated simulations ($\beta_{0}^{\rm mid} \lesssim 10^{2}$).

\section{Discussion}
\label{sec:disc}

\subsection{MRI-Dynamo Cycles}
We find that the cyclic MRI-dynamo reversals in the large-scale toroidal magnetic field have a regular $\simeq 10$ orbit periodicity for simulations with zero and weak ($\beta_{0}^{\rm mid} \gtrsim 10^{3}$) net vertical magnetic flux.  For $\beta_{0}^{\rm mid} \lesssim 10^{3}$, these reversals become more scattered and increase in period with increasing net vertical magnetic flux, until they are no longer observed for the most strongly magnetized simulation with $\beta_{0}^{\rm mid} = 10^{1}$.

Previous shearing box simulations with a zero net vertical magnetic flux show the same MRI-dynamo cycle regularity and periodicity that we observe, provided that the domain size is sufficiently large \citep[e.g.,][]{Simon2012}.  The strong ($\beta_{0}^{\rm mid} \lesssim 10^{3}$) net vertical magnetic flux simulations of \citet{BaiStone2013} also display MRI-dynamo cycle characteristics that are consistent with our simulations.  However, some differences from our results arise for weaker net fluxes, with previous works finding the butterfly pattern to be highly irregular for $\beta_{0}^{\rm mid} \gtrsim 10^{4}$ \citep[e.g.,][]{Fromang2013, BaiStone2013}. Given that \citet{BaiStone2013} used very similar methods to ours, the most likely origin for the differences we observe is differences in the domain size. The shearing box simulations of \citet{BaiStone2013} used a substantially smaller horizontal domain ($L_{x}, L_{y}) = (4 H_{0}, 8 H_{0}$), which has been shown in zero net vertical magnetic flux simulations to affect the dynamo properties \citep{Simon2012}.  

The smooth transition that we observe in the cyclic MRI-dynamo behaviour across the zero-to-weak net vertical magnetic flux regimes suggests that we have obtained numerically converged solutions, that are physically reasonable in the local limit. However, the requirement for such large horizontal domains means that we are pushing the limits of the shearing box's validity by neglecting curvature terms.  We cannot rule out the possibility that the MRI-dynamo is intrinsically periodic in a shearing box and even the strong field models may become periodic for larger domain sizes.  Similarly, the absence of MRI-dynamo cycles we observe for $\beta_{0}^{\rm mid} = 10^{1}$ may be a consequence of an insufficiently small domain size and/or short integration time.  Ultimately, global disc simulations are needed to determine if the sporadic nature of MRI-dynamo cycles that we see in the strong ($\beta_{0}^{\rm mid} \lesssim 10^{3}$) net vertical magnetic flux regime is an intrinsic property of MRI-dynamo behaviour.

\subsection{Quasi-Periodic Oscillations}
\label{sec:disc_QPOs}
Black hole X-ray binaries showcase complicated evolutionary cycles, called state transitions \citep[e.g.,][]{RemillardMcClintock2006, Belloni2010}, where the source transitions between ``low/hard'' and ``high/soft'' states.  The low/hard state is characterized by a low luminosity ($\lesssim 10^{-3} L_{\rm Edd}$) and a non-thermal hard X-ray spectrum, while the high/soft state is characterized by a high luminosity ($\gtrsim 0.1 L_{\rm Edd}$) and a quasi-thermal soft X-ray spectrum.  The soft quasi-thermal X-rays are attributed to a geometrically thin, optically thick accretion disc \citep{ShakuraSunyaev1973}, while the non-thermal hard X-rays are thought to arise from energetic electrons in a surrounding ``corona'' that inverse Compton scatter the seed disc photons \citep{HaardtMaraschi1991}.  In a complete state transition hysteresis cycle ($\sim 100$ days), the source rises out of the low/hard state along a high-luminosity track into the high/soft state, but then decays on a different low-luminosity track back to the low/hard state, passing through hybrid ``intermediate'' states along the way.  The outburst decay track is typically $\sim10-100$ times fainter than the outburst rise track.

During a low/hard $\leftrightarrow$ high/soft state transition, low-frequency quasi-periodic oscillations (QPOs) are observed \citep{vanderKlis1989, Casella2005}, which are coherent features in the temporal power spectrum of the X-ray light curve.  The QPO centroid frequency evolves from $\nu_{{\rm Q}, 0} \simeq 0.01 \rightarrow 10~{\rm Hz}$ during a low/hard $\rightarrow$ high/soft state transition \citep[e.g.,][]{Belloni2005, Belloni2010}.  These QPOs ultimately disappear in the high/soft state.  The origin of QPOs is still unknown, but if understood, QPOs would be a powerful diagnostic for measuring disc evolution \citep[e.g.,][]{MillerMiller2014}.

The MRI-dynamo cycles of flip-flopping toroidal magnetic field shown in Figure \ref{fig:butterfly} are promising candidates for QPOs \citep[e.g.,][]{ONeill2011, Begelman2015}.  Figure \ref{fig:powspec_By} shows that a coherent QPO-like feature appears for the weakly magnetized discs ($\beta_{0}^{\rm mid} \gtrsim 10^{4}$) with peak frequency $\nu_{\rm peak} \sim 0.1~{\rm orbits^{-1}}$.  A strong peak is not seen for the strongly magnetized discs ($\beta_{0}^{\rm mid} \lesssim 10^{3}$) due to the irregularity of the MRI-dynamo cycles, but this result may be a consequence of domain size restrictions as previously mentioned.  We consider the increase in MRI-dynamo period with net flux to be a robust result; therefore, the frequency of a resulting power spectral peak would increase with decreasing magnetization.  However, we emphasize that Figure \ref{fig:powspec_By} cannot be directly compared to an observed power spectrum because the toroidal magnetic field was not mapped to an emission mechanism.  If, as suggested by \citet{BegelmanArmitage2014}, the disc becomes decreasingly magnetized during the low/hard $\rightarrow$ high/soft state transition, MRI-dynamo cycles could potentially produce a QPO-like feature that fits the observed evolutionary behaviour of $\nu_{\rm Q}$.  Global simulations are needed to address these speculations \citep[e.g.,][]{ONeill2011} because QPO frequency and strength would likely be set far from the black hole, and the modulations would have to take into account radial propagation of the MRI-dynamo cycles.

\subsection{Spectral Hardening}
\label{sec:disc_bhspin}
Understanding the consequences of vertical magnetic pressure support for the observed accretion disc spectrum is necessary for assessing the robustness of any observational results based on disc continuum modeling.  A disc with magnetic pressure support has a lower gas density at the effective photosphere, which enhances electron scattering and leads to a harder spectrum compared to a disc without magnetic pressure support \citep{Davis2005, Blaes2006}.

The degree of spectral hardening is parametrized by the colour correction factor $f_{\rm col}$ \citep{ShimuraTakahara1995b}.  Discs without a vertical magnetic pressure contribution tend to have $f_{\rm col}$ values confined to the narrow range $f_{\rm col} \simeq 1.5 - 1.7$ \citep[e.g.,][]{Davis2005}.  However, observations of disc evolution in X-ray binary state transitions \citep{Salvesen2013, ReynoldsMiller2013} and sophisticated joint-spectral fitting of the disc in a steady high/soft state \citep{Maitra2014} demonstrate that the data permit $f_{\rm col} \gtrsim 2$.  Post-processing of a weakly magnetized shearing box simulation with zero net vertical magnetic flux showed that magnetic pressure support in the disc surface layers leads to enhanced spectral hardening, with $f_{\rm col} = 1.74$ (1.48) when magnetic pressure support was included (neglected) in the vertical hydrostatic balance \citep{Blaes2006}.  Analytic models of very strongly magnetized discs predict substantial spectral hardening of $f_{\rm col} \sim 5$ \citep{BegelmanPringle2007}.  The two main contributing effects to spectral hardening are the ratio of electron scattering to absorption being higher at the disc photosphere and the profile of MRI dissipation being concentrated higher in the disc (i.e., there is a lot of dissipation just below the photosphere).  However, the quantitative details of how significant magnetic pressure support influences the observed disc spectrum are not yet understood.

Measurements of black hole spin with the disc continuum fitting technique \citep{Zhang1997, McClintock2006} ignore vertical magnetic pressure support in the disc when modeling the disc spectrum.  This may lead to erroneously overestimated spins and explain the inconsistencies between the two leading black hole spin measurement techniques.  Given a non-spinning black hole ($a = 0$) with typical X-ray binary system parameters, one would obtain the incorrect spin parameter $a \simeq 0.5$ by ignoring fairly weak vertical magnetic pressure support \citep{Blaes2006}.  We are currently quantifying the impact of spectral hardening on black hole spin measurements (Salvesen et al. 2016, in preparation).

\section{Summary and Conclusions}
\label{sec:sumconc}
We have studied the structure and dynamo-related variability of accretion discs, using isothermal local simulations that include vertical stratification. The simulations span almost the full range of net vertical fluxes that lead to MRI-unstable initial conditions ($\beta_{0}^{\rm mid} = 10^{1} - 10^{5}$), and were run in large domains that capture mesoscale structures that are important for disc variability. Our two strongest net field runs, with $\beta_{0}^{\rm mid} = 10$ and 100, result in the formation of magnetically dominated discs whose structure differs qualitatively from that of gas pressure-dominated solutions. Our main conclusions are:
\begin{itemize}
\item 
The $\alpha$-viscosity parameter, $\overline{\alpha} = \langle T_{xy} \rangle_{V} / \langle p_{\rm gas} \rangle_{V}$, follows a power-law spanning two orders of magnitude in net vertical magnetic flux (four orders of magnitude in $\beta_{0}^{\rm mid}$). Alternatively, normalizing the stress to the magnetic pressure yields a constant viscosity parameter $\overline{\alpha}_{\rm mag} = \langle T_{xy} \rangle_{V} / \langle p_{B} \rangle_{V}$ that is independent of the initial mid-plane $\beta$ \citep{Hawley1995, Blackman2008}.
\item 
Gas density fluctuations (relative to the time-averaged gas density vertical profile) increase with net vertical magnetic flux, reaching the $\sim 60-100\%$ level for our magnetic-pressure dominated disc simulations NVF-$\beta 2$ and NVF-$\beta 1$.  These highly magnetized but still MRI-unstable discs are highly inhomogeneous, clumpy structures.
\item 
MRI-dynamo cycles are highly regular for simulations with either zero net vertical magnetic flux or weak ($\beta_{0}^{\rm mid} \gtrsim 10^{4}$) net vertical magnetic flux.  The MRI-dynamo period increases and becomes more sporadic for moderate-to-strong ($\beta_{0}^{\rm mid} \lesssim 10^{3}$) net vertical magnetic flux simulations. A large horizontal domain appears to be necessary to capture this behaviour, in net field as well as zero net field simulations \citep{Simon2012}.
\item
Entry into the fully magnetic pressure-dominated regime is not a sufficient condition for the complete suppression of cyclic MRI-dynamo reversals. We observe multiple reversals in the magnetic pressure-dominated NVF-$\beta 2$ simulation, and cannot exclude the possibility that very long time scale reversals occur for even more strongly magnetized discs.  With increasing magnetization, even longer integration times and/or larger domain sizes than we considered may be necessary to properly study these MRI-dynamo cycles.
\item 
The magnetization at the disc mid-plane is very well-described by the regularity condition (see Equation \ref{eqn:beta_B15}) given by \citet{Begelman2015}, which is derived from a model based on the balance between the MRI-dynamo production of toroidal field and its buoyant escape.
\item
The \citet{Begelman2015} model provides a reasonable first-order description of the vertical disc structure seen in the simulation results.  Incorporating a magnetic reconnection efficiency parameter $\nu$ into the model helps to match the steep magnetic pressure profiles from the magnetic pressure-dominated disc simulations ($\beta_{0}^{\rm mid} \lesssim 10^{2}$).
\end{itemize}

Our results support aspects of a currently speculative scenario in which the net poloidal flux is the key parameter controlling the structure and evolution of accretion discs in black hole X-ray binaries. Net fluxes that remain in the MRI-unstable regime can still be strong enough to stimulate qualitative changes to the disc structure, and lead to the formation of magnetically supported and highly inhomogeneous discs. Analytic models of such discs suggest that many of the observationally interesting consequences occur as a consequence of the work done by the toroidal field as it escapes, and future simulations will need to relax the isothermal assumption adopted here to study such effects.

\section*{Acknowledgments}
We thank the anonymous referee for her/his constructive comments and suggestions, which improved this paper. GS acknowledges support through the NASA Earth and Space Science Graduate Fellowship program.  J.B.S.'s support was provided in part under contract with the California Institute of Technology (Caltech) and the Jet Propulsion Laboratory (JPL) funded by NASA through the Sagan Fellowship Program executed by the NASA Exoplanet Science Institute. PJA acknowledges support from NASA under Astrophysics Theory Program awards NNX11AE12G and NNX14AB42G, and from the NSF under award AST-1313021. MCB acknowledges support from NSF grant AST-1411879.  This work used the \texttt{Janus} supercomputer, which is supported by the National Science Foundation (award number CNS-0821794) and the University of Colorado Boulder.  The \texttt{Janus} supercomputer is a joint effort of the University of Colorado Boulder, the University of Colorado Denver, and the National Center for Atmospheric Research.  This work used the \texttt{yt} project \citep{Turk2011}, an open source data analysis and visualization toolkit for astrophysical simulations.

\bibliographystyle{mn2e}

\label{lastpage}
\end{document}